\documentclass[twocolumn,showpacs,aps,epsfig,nofootinbib]{revtex4}
\usepackage{graphicx}
\usepackage{amsfonts}
\usepackage{epstopdf}
\usepackage{latexsym}
\usepackage{amssymb}
\usepackage{amssymb}

\usepackage[center]{subfigure}

\begin{document}

%%%%%%%%%%%%%%%%%%%%%%%%%%%%%%%%%%%%%%%%%%%%%%%%%%%%%%%%%%%%%%%
 \newcommand{\bq}{\begin{equation}}
 \newcommand{\eq}{\end{equation}}
 \newcommand{\bqn}{\begin{eqnarray}}
 \newcommand{\eqn}{\end{eqnarray}}
 \newcommand{\nb}{\nonumber}
 \newcommand{\lb}{\label}
\newcommand{\PRL}{Phys. Rev. Lett.}
\newcommand{\PL}{Phys. Lett.}
\newcommand{\PR}{Phys. Rev.}
\newcommand{\CQG}{Class. Quantum Grav.}
 %%%%%%%%%%%%%%%%%%%%%%%%%%%%%%%%%%%%%%%%%%%%%%%%%%%%%%%%%%%%%%%

\begin{flushright}

{\mbox{\hspace{10cm}}} IPMU12-0115

\end{flushright}

\title{Solar system tests and interpretation of gauge field and
Newtonian prepotential in general covariant Ho\v{r}ava-Lifshitz gravity}

\author{Kai Lin ${}^{a}$}
\email{K_Lin@baylor.edu}

\author{Shinji Mukohyama ${}^{b}$}
\email{shinji.mukohyama@ipmu.jp}

\author{Anzhong Wang ${}^{a, c}$}
\email{anzhong_wang@baylor.edu}

\affiliation{ ${}^{a}$ Institute  for Advanced Physics $\&$ Mathematics,   Zhejiang University of
Technology, Hangzhou 310032,  China \\
${}^{b}$  Kavli Institute for the Physics and Mathematics of the 
Universe (Kavli IPMU), TODIAS,  the University of Tokyo,\\ 5-1-5
Kashiwanoha, Kashiwa, 277-8583, Japan \\
${}^{c}$ GCAP-CASPER, Physics Department, Baylor
University, Waco, TX 76798-7316, USA }
{}

\date{\today}

\begin{abstract}
 In this paper, we first study spherically symmetric, stationary vacuum
 configurations in general covariant theory ($U(1)$ extension) of
 Ho\v{r}ava-Lifshitz gravity with the projectability condition and an 
 arbitrary value of the coupling constant $\lambda$. We obtain all the
 solutions with the assumed symmetry in closed forms. If the gauge field
 $A$ and the Newtonian prepotential $\varphi$ do not directly couple to 
 matter fields, the theory is inconsistent with solar system tests for 
 $\lambda\not=1$, no matter how small $|\lambda-1|$ is. This is shown
 to be true also with the most general ansatz of spherically symmetric
 (but not necessarily stationary) configurations. Therefore, to be
 consistent with observations, one needs either to find a mechanism to
 restrict $\lambda$ precisely to its relativistic value
 $\lambda_{GR}=1$, or to consider $A$ and/or $\varphi$ as parts of the
 $4$-dimensional metric on which matter fields propagate. In the latter,
 requiring that the line element be invariant not only under the
 foliation-preserving diffeomorphism but also under the local $U(1)$
 transformations, we propose the replacements, 
 $N \rightarrow N - \upsilon(A - {\cal{A}})/c^2$ and 
 $N^i \rightarrow N^i+N\nabla^{i}\varphi$, where $\upsilon$ is a 
 dimensionless coupling constant to be constrained by observations, $N$
 and $N^i$ are, respectively, the lapse function and  shift vector,
 and ${\cal{A}} \equiv - \dot{\varphi} + N^i\nabla_{i}\varphi +
 N(\nabla_{i}\varphi)^2/2$. With this prescription, we show explicitly
 that the aforementioned solutions are consistent with solar system
 tests for both $\lambda=1$ and $\lambda\not=1$, provided that
 $|\upsilon-1|<10^{-5}$. From this result, the physical and geometrical 
 interpretations of the fields $A$  and
 $\varphi$ become clear. However, it still remains to be understood how 
 to obtain such a prescription from the action principle.   
 
\end{abstract}

\pacs{04.60.-m; 98.80.Cq; 98.80.-k; 98.80.Bp}

\maketitle

% \tableofcontents

\section{Introduction}
\renewcommand{\theequation}{1.\arabic{equation}} \setcounter{equation}{0}

Einstein's classical general theory of relativity (GR) is consistent
with all the experiments and observations carried out so far
\cite{Will05}. However, it has been known for a long time that GR is not
(perturbatively) renormalizable \cite{HV}, and thus can be considered 
only as a low energy effective theory. Because of the universal coupling
of gravity to all forms of energy, it is expected that gravity too
should have a quantum mechanical description. Motivated by this strong
anticipation, quantization of gravitational fields has been one of the
main driving forces in theoretical physics in the past decades in a wide
range of approaches \cite{QGs}.

Recently, Ho\v{r}ava \cite{Horava} proposed a new theory of quantum
gravity in the framework of quantum field theory. One of the essential
ingredients of the theory is the inclusion of higher-dimensional spatial
(but not time) derivative operators, so that the ultraviolet (UV) behavior is
dominated by them and that they render the theory power-counting
renormalizable. In the infrared (IR) the lower dimensional operators
take over, presumably providing a healthy low energy limit. The
exclusion of higher time derivative terms prevents ghost instability
\cite{Stelle}, but breaks Lorentz symmetry, on the other hand. While the
breaking of Lorentz symmetry in the matter sector is highly restricted
by experiments/observations, in the gravitational sector the
restrictions are much weaker \cite{LZbreaking} (See also
\cite{Pola}). The Lorentz breaking and hence the power-counting
renormalizability are realized by invoking the anisotropic scaling
between time and space, 
\bq
\lb{1.1}
t \rightarrow b^{-z} t,\;\;\; \vec{x} \rightarrow b^{-1}\vec{x}.
\eq
This is a reminiscent of Lifshitz scalars \cite{Lifshitz} in condensed
matter physics, hence the theory is often referred to as the
Ho\v{r}ava-Lifshitz (HL) gravity. For the theory to be power-counting
renormalizable, the critical exponent $z$ has to be $z \ge 3$
\cite{Horava,Visser}.  Clearly, such a scaling breaks explicitly the
Lorentz symmetry and thus $4$-dimensional diffeomorphism
invariance. Ho\v{r}ava assumed that it is broken only down to, the
invariance under 
\begin{equation}
\lb{1.2}
 t \to t'(t), \quad \vec{x}\to\vec{x}'(t,\vec{x}),
\end{equation}
the so-called foliation-preserving diffeomorphism,  denoted often by
Diff($M, \; {\cal{F}}$). The basic quantities are the lapse function
$N$, the shift vector $N^i$, and the $3$-dimensional spatial metric
$g_{ij}$, as introduced more than 50 years ago by Arnowitt, Deser and
Misner \cite{ADM}, in order to quantize gravity.

Once the general covariance is broken, it immediately results in a 
proliferation of independent coupling constants
\cite{Horava,BPS,KP,ZSWW}, which could potentially limit the 
predictive power of the theory. To reduce the number of independent
coupling constants,  Ho\v{r}ava introduced two independent 
conditions, the {\em projectability} and the {\em detailed balance}
\cite{Horava}. The former requires that the lapse function $N$ be a
function of $t$ only, 
\bq
\lb{1.3}
N = N(t),
\eq
while the latter requires that
the gravitational potential  should be obtained from a superpotential $W_{g}$,
where  $W_{g}$  is given by an integral of
the gravitational Chern-Simons term over a 3-dimensional space, 
$W_{g}\sim \int_{\Sigma}{\omega_{3}(\Gamma)}$. 
With these two conditions, the general action contains only five
independent  coupling constants. 
The detailed balance condition has several remarkable features
\cite{Horava,BLW,ZSWW}.  For example,  it is in the same spirit of the
AdS/CFT correspondence \cite{AdSCFT}, where a string theory including
gravity defined on one space is equivalent to a quantum field theory
without gravity defined on the conformal boundary of this space, which
has one or more lower dimension(s).  Yet, in the non-equilibrium
thermodynamics, the counterpart of the superpotential $W_{g}$  plays the
role of entropy, while ${\delta{W}_{g}}/{\delta{g}_{ij}}$ represents
the corresponding entropic force \cite{OM}. This might shed further
lights on the nature of gravitational forces, as proposed recently by
Verlinde \cite{Verlinde}. For details, we refer readers to  Ho\v{r}ava's
original paper \cite{Horava}, as well as his review article \cite{Hreview}.

When applying the theory to cosmology, various remarkable features were
found. (See \cite{Mukohyama:2010xz} for a review.) In particular, the
higher-order spatial curvature terms can give rise to a bouncing
universe \cite{Calcagni}, may ameliorate the flatness problem \cite{KK} 
and lead to caustic avoidance \cite{Mukohyama:2009tp}; the anisotropic
scaling provides a solution to the horizon problem and generation of
scale-invariant perturbations without inflation \cite{Mukohyama:2009gg},
a new mechanism for generation of primordial magnetic seed field
\cite{MMS}, and also a modification of the spectrum of gravitational 
wave background via a peculiar scaling of radiation energy density
\cite{MNTY}; with the projectability condition, the lack of a local
Hamiltonian constraint leads to ``dark matter as an integration
constant'' \cite{Mukohyama:2009mz}; the dark sector can also have its
purely geometric origins \cite{Wang}; in the parity-violating version of
the theory, circularly polarized gravitational waves can also be
generated in the early universe \cite{TS}; and so on.

Despite of all the above remarkable features, it was found that the
projectability condition leads to several undesirable properties,
including infrared instability \cite{Horava,Ins} and strong coupling
\cite{KA,WWa} \footnote{Note that even without the 
projectability condition the theory is still strongly coupled
\cite{SC,KP}, although instability can be avoided by inclusion of the
term $a_ia^i$ \cite{BPS}, where $a_{i} = N_{,i}/N$. On the other hand,
as mentioned above, abandoning the projectability condition results in a
proliferation of independent coupling constants. To render this problem,
Zhu, Wu, Wang, and Shu recently introduced a local U(1) symmetry (See
Eq.(\ref{symmetry}) given below), in addition to the detailed balance
condition \cite{ZWWS}. In order to have a healthy IR limit, however,
they found that the latter has to be broken softly by adding all the low 
dimensional relevant terms. Even with these terms, the number of the
independently coupling constants is reduced to $15$.}. 
All these properties are closely related to the existence of a spin-0
graviton \cite{reviews,Mukohyama:2010xz}.

It should be noted, however, that the infrared instability does not show
up under a certain condition \cite{Mukohyama:2010xz} and that the strong
coupling is not necessarily a problem if nonlinear effects help
recovering GR  at low energy. Of course, the strong
coupling implies that the naive perturbative expansion breaks down and
that a proper non-perturbative treatment is needed. In general,
non-perturbative analysis is not easy to perform in
practice. Nonetheless, in some simplified situations, fully nonlinear
analysises were already performed, showing that the $\lambda\to 1$ limit
of the theory is continuous and that GR is recovered in a
non-perturbative fashion. Such examples include spherically symmetric,
stationary, vacuum configurations \cite{Mukohyama:2010xz}, a class of
exact cosmological solutions \cite{WWa} and nonlinear superhorizon
perturbations \cite{Izumi:2011eh,GSWa}. The non-perturbative recovery of
GR, explicitly shown in those examples, may be considered as an analogue
of the Vainshtein effect \cite{Vainshtein:1972sx}.

Although the existence of spin-0 graviton after all may not be a
problem due to the analogue of the Vainshtein effect, it is interesting
and certainly important to seek another possible way out. Motivated by
this, Ho\v{r}ava and Melby-Thompson (HMT)  \cite{HMT} extended the
symmetry (\ref{1.2}) to include a local $U(1)$,  
\bq
\lb{symmetry}
 U(1) \ltimes {\mbox{Diff}}(M, \; {\cal{F}}).
\eq
With this enlarged symmetry, the spin-0 graviton is eliminated 
\cite{HMT,WW}, and the theory has the same number of propagating degrees
of freedom as GR. This was initially done in the special case with
$\lambda=1$, and was soon generalized to the case with any 
$\lambda$ \cite{Silva}. Even with $\lambda\ne 1$, the spin-0 graviton
is still eliminated \cite{Silva,HW}. When applying it to cosmology, various interesting
results were found \cite{HWWb}. In particular,  the
Friedmann-Robterson-Walker (FRW) universe is necessarily flat in such a
setup, provided that the coupling of the U(1) field to a scalar matter
field is described by the recipe given in \cite{Silva}.

In this paper, we shall consider two important issues in the general
covariant theory of the HL gravity with the projectability condition
(\ref{1.3}) and an arbitrary coupling constant $\lambda$
\cite{HMT,Silva,HW}: (i) the solar system tests; and (ii) the physical
and geometrical interpretations of the gauge field $A$ and Newtonian 
pre-potential $\varphi$. Specifically, after giving a brief introduction
to the theory in Sec. II, we present all the spherically symmetric,
stationary, vacuum solutions of the theory in closed forms in
Sec. III. In Sec. IV, we consider the solar system tests by not taking
$A$ and $\varphi$ as parts of the low energy $4$-dimensional metric on
which matter fields propagate, and find that theory is not consistent
with observations as long as $\lambda$ is not precisely equal to one,
however small $|\lambda-1|$ is. In Sec. V, we further study the  
limit $\lambda \rightarrow 1$ without assuming that the configuration is
stationary. We find that, although the limit exists, it does not reduces
to the Schwarzschild geometry. These resuts in Sec. IV and V strongly
suggest that, in order for the theory to be consistent with the solar
system tests, $A$ and/or $\varphi$ should enter the low-energy
$4$-dimensional metric. Thus, in Sec. VI, by requiring that the line
element $ds^2$ be gauge-invariant not only under Diff($M, {\cal{F}}$),
but also under the $U(1)$ transformations, we propose that it should
take the form, 
\bq
\lb{line}
ds^{2} = - {\cal{N}}^2 c^2 dt^2 + g_{ij}\left(dx^i + {\cal{N}}^idt\right) \left(dx^j + {\cal{N}}^jdt\right),
\eq
where
\bqn
\lb{ident}
{\cal{N}} &\equiv& N - \frac{\upsilon}{c^2}(A - {\cal{A}}),\;\;\; {\cal{N}}^i \equiv  N^i + N\nabla^i\varphi,\nb\\
{\cal{A}} &\equiv& - \dot{\varphi} + N^i\nabla_{i}\varphi + 
\frac{1}{2}N(\nabla_{i}\varphi)^2,
\eqn
where $\upsilon$ is a dimensionless coupling constant to be constrained
by experiments/observations, and subjected to radiative
corrections. $\nabla_{i}$ denotes the covariant derivative with respect to the 3-metric $g_{ij}$. With such replacements, in this section we show explicitly
that the resulted metrics are consistent with observations for both
$\lambda = 1$ and $\lambda \not=1$.  With these replacements, one also 
sees clearly the physical and geometric meanings of $A$ and
$\varphi$. Our main results are summarized and discussed in Sec. VII. 

Note that solar system tests were studied in other versions of the HL theory previously
\cite{SST}. However, to our best knowledge, in the current paper it is the first time to consider the problem in the general covariant
theory of the HL gravity with the projectability condition and an arbitrary coupling constant $\lambda$,
while the case with $\lambda = 1$ was studied in \cite{GSW}.

In addition, all the high-order derivative terms of curvatures are negligible  in the IR. Then, 
 test particles move along geodesics, as shown explicitly in \cite{GLLSW,BLWb} by using optical 
 geometric approximations. Therefore, to have a consistent treatment, 
 when we consider solar system tests, we ignore all the 
 corrections from these high-order terms.

\section{General covariant HL theory }%with any given $\lambda$}
\renewcommand{\theequation}{2.\arabic{equation}} \setcounter{equation}{0}

To realize the enlarged symmetry (\ref{symmetry}), HMT observed that
the linearized (minimal) HL theory has a global U(1) symmetry for
$\lambda = 1$. This symmetry can be promoted to a local one by
introducing a gauge field $A$, with which it was found that the scalar
degree of freedom is eliminated \cite{HMT}. When they lifted it to a
full nonlinear theory, HMT found that the realization of the symmetry
(\ref{symmetry}) requires introduction of an auxiliary scalar field
$\varphi$, which was referred to as the  ``Newtonian prepotential.''
Under the local U(1), both $A$ and $\varphi$ transform as, 
\bq
\lb{2.0a}
\delta_\alpha A=\dot{\alpha}-N^i\nabla_i\alpha,\;\;\;
\delta_\alpha\varphi= - \alpha,
\eq
while the lapse function $N$, the shift vector $N^i$ and the 3-metric
$g_{ij}$  transform as,
\bqn
\lb{gauge}
\delta_\alpha N = 0,\;\;\; \delta_\alpha N_i=N\nabla_i\alpha,\;\;\; \delta_\alpha g_{ij}=0,
\eqn
where $\alpha$ denotes the U(1) generator, and 
$\dot{\alpha} \equiv \partial\alpha/\partial t$.

Under the coordinate transformations (\ref{1.2}), $\varphi$ transforms as a scalar, while $A$ transforms as a vector under the time reparametrizations 
$ t \rightarrow  f(t')$, and as a scalar under the spatial transformations $\vec{x}\to\vec{\zeta}(t',\vec{x}')$, namely, 
\bqn
\lb{2.0b}
\delta{A} &=& \zeta^{i}\partial_{i}A + \dot{f}A  + f\dot{A},\nb\\
\delta \varphi &=&  f \dot{\varphi} + \zeta^{i}\partial_{i}\varphi.
\eqn
The metric components, $N,\; N^i$ and $g_{ij}$,  on the other hand,  transform as
\bqn
\lb{1.5}
\delta{N} &=& \zeta^{k}\nabla_{k}N + \dot{N}f + N\dot{f},\nb\\
\delta{N}_{i} &=& N_{k}\nabla_{i}\zeta^{k} + \zeta^{k}\nabla_{k}N_{i}  + g_{ik}\dot{\zeta}^{k}
+ \dot{N}_{i}f + N_{i}\dot{f}, \nb\\
\delta{g}_{ij} &=& \nabla_{i}\zeta_{j} + \nabla_{j}\zeta_{i} + f\dot{g}_{ij}, 
\eqn
under  (\ref{1.2}).

The HMT model was initially constructed in the case $\lambda = 1$, and it was soon found that it can be generalized 
to the case with an arbitrary $\lambda$ \cite{Silva}, in which the spin-0 gravitons are also eliminated \cite{Silva,HW}, so the gravitational sector
has the same degree of freedom as that in GR, i.e., only massless spin-2 gravitons exist. 

For any given coupling constant $\lambda$, the total action can be
written as \cite{HMT,WW,Silva,HW}, 
 \bqn \lb{2.4}
S &=& \zeta^2\int dt d^{3}x N \sqrt{g} \Big({\cal{L}}_{K} -
{\cal{L}}_{{V}} +  {\cal{L}}_{{\varphi}} +  {\cal{L}}_{{A}} +  {\cal{L}}_{{\lambda}} \nb\\
& & ~~~~~~~~~~~~~~~~~~~~~~ \left. + {\zeta^{-2}} {\cal{L}}_{M} \right),
 \eqn
where $g={\rm det}\,g_{ij}$, and
 \bqn \lb{2.5}
{\cal{L}}_{K} &=& K_{ij}K^{ij} -   \lambda K^{2},\nb\\
{\cal{L}}_{\varphi} &=&\varphi {\cal{G}}^{ij} \Big(2K_{ij} + \nabla_{i}\nabla_{j}\varphi\Big),\nb\\
{\cal{L}}_{A} &=&\frac{A}{N}\Big(2\Lambda_{g} - R\Big),\nb\\
{\cal{L}}_{\lambda} &=& \big(1-\lambda\big)\Big[\big(\nabla^{2}\varphi\big)^{2} + 2 K \nabla^{2}\varphi\Big].
 \eqn
Here %$ \lambda \equiv 1 - \xi,\;
$\Lambda_{g}$ is a    coupling constant, and the
Ricci and Riemann terms all refer to the three-metric $g_{ij}$, and
 \bqn \lb{2.6}
K_{ij} &=& \frac{1}{2N}\left(- \dot{g}_{ij} + \nabla_{i}N_{j} +
\nabla_{j}N_{i}\right),\nb\\
{\cal{G}}_{ij} &=& R_{ij} - \frac{1}{2}g_{ij}R + \Lambda_{g} g_{ij}.
 \eqn
% where %$N_{i} = g_{ij}N^{j}$.
${\cal{L}}_{M}$ is the
matter Lagrangian density, which in general is a function of all the dynamical variables,
$U(1)$ gauge field, and the Newtonian prepotential, i.e.,
$
{\cal{L}}_{M} = {\cal{L}}_{M}\big(N, \; N_{i}, \; g_{ij}, \; \varphi,\; A; \; \chi\big)$,
%\eq
where $\chi$ denotes collectively the matter fields. ${\cal{L}}_{{V}}$ is an arbitrary Diff($\Sigma$)-invariant local scalar functional
built out of the spatial metric, its Riemann tensor and spatial covariant derivatives, without the use of time derivatives.

Note the difference between the notations used here and the ones used in \cite{HMT,Silva} \footnote{In particular, we
have $K_{ij} = - K_{ij}^{HMT},\; \Lambda_{g} = \Omega^{HMT},\; \varphi = - \nu^{HMT}, {\cal{G}}_{ij} = \Theta_{ij}^{HMT}$,
where quantities with the super-indice ``HMT" are those used in \cite{HMT,Silva}.}. In this paper, without further explanations,
we shall use directly  the notations and conventions defined in \cite{WM} and \cite{WW}.
%, which will be referred, respectively, to as Paper I and Paper II.
%However, in order to have the present paper as independent as possible, it is difficult to avoid
%repeating the same materials, although we shall try to limit it to its minimum.

In \cite{SVW}, by assuming that the highest order derivatives are six, the minimum in order to have the theory  to be power-counting
renormalizable \cite{Horava,Visser}, and that  the theory  preserves
the parity,  the most general form of  ${\cal{L}}_{{V}}$ was constructed and is given by, 
 \bqn \lb{2.5a}
{\cal{L}}_{{V}} &=& \zeta^{2}g_{0}  + g_{1} R + \frac{1}{\zeta^{2}}
\left(g_{2}R^{2} +  g_{3}  R_{ij}R^{ij}\right)\nb\\
& & + \frac{1}{\zeta^{4}} \left(g_{4}R^{3} +  g_{5}  R\;
R_{ij}R^{ij}
+   g_{6}  R^{i}_{j} R^{j}_{k} R^{k}_{i} \right)\nb\\
& & + \frac{1}{\zeta^{4}} \Big[g_{7}(\nabla R)^{2} +  g_{8}
\left(\nabla_{i}R_{jk}\right) \left(\nabla^{i}R^{jk}\right)\Big],
~~~~
 \eqn
 where the coupling  constants $ g_{s}\, (s=0, 1, 2,\dots 8)$  are all dimensionless, and
 \bq
 \lb{Lambda}
 \Lambda = \frac{1}{2} \zeta^{2}g_{0},
 \eq
 is the cosmological constant. The relativistic limit in the IR
 requires
 \bq
 \lb{GR}
 g_{1} = -1,\;\;\; \zeta^2 = \frac{1}{16\pi G}.
 \eq

Then, the corresponding field equations are given in Appendix A.

\section{Spherical Vacuum Solutions }
\renewcommand{\theequation}{3.\arabic{equation}} \setcounter{equation}{0}

Spherically symmetric static vacuum spacetimes with projectability condition in the HMT setup were  studied systematically in \cite{GSW,GLLSW,AP,BLWb}.
In particular, the ADM quantities   can be cast in the form \cite{GPW,IM},
\bqn
\lb{4.1}
&& N = 1, \;\;\; N^i\partial_{i} = e^{\mu-\nu} \partial_{r},\nb\\
&& g_{ij}dx^idx^j =   e^{2\nu} dr^2  + r^{2}d^2\Omega,
\eqn
in the spherical coordinates $x^{i} = (r, \theta, \phi)$, where  $d^2\Omega = d\theta^{2}  + \sin^{2}\theta d\phi^{2}$,
and
\bq
\lb{4.2}
 \mu = \mu(r),\;\;\; \nu = \nu(r). %\;\;\; N^{i} = e^{\mu - \nu}\delta^{i}_{r}.
 \eq
The corresponding timelike Killing vector is  $\xi = \partial_{t}$. In the diagonal case, we have $\mu = -\infty$. 
With the gauge freedom of the local U(1) symmetry,  without loss of the generality, we  can always fix the gauge by setting
 \bq
 \lb{gauge}
 \varphi =0.
 \eq
 Then,   we find that
 \bq
 \lb{4.2a}
 F_{\varphi}^{ij} = 0,\;\;\; F_{\varphi}^{i} = 0,\;\;\; {\cal{L}}_{\varphi} = {\cal{L}}_{\lambda} = 0,
 \eq
 and
\bqn
\lb{4.3}
K_{ij} &=& e^{\mu+\nu}\Big(\mu'\delta^{r}_{i}\delta^{r}_{j} + re^{-2\nu}\Omega_{ij}\Big),\nb\\
R_{ij} &=&  \frac{2\nu'}{r}\delta^{r}_{i}\delta^{r}_{j} + e^{-2\nu}\Big[r\nu' - \big(1-e^{2\nu}\big)\Big]\Omega_{ij},\nb\\
%{\cal{L}}_{\varphi} &=& 0,\;\;\; F_{\varphi}^{ij} = 0,\nb\\
\pi_{ij} &=&\frac{e^{\mu+\nu}}{r}\big[2\lambda+(\lambda-1)r\mu'\big]\delta_i^r\delta_j^r\nb\\
& &+re^{\mu-\nu}\big[2\lambda-1+\lambda r\mu'\big]\Omega_{ij}\nb\\
{\cal{L}}_{K} &=& - \frac{e^{2(\mu-\nu)}}{r^{2}} \left[4\lambda-2+4\lambda r\mu' + r^2(\lambda-1)(\mu')^2\right],\nb\\
{\cal{L}}_{A} &=&  \frac{2A}{r^2} \Big[e^{-2\nu}\left(1 - 2r
\nu'\right) + \left(\Lambda_{g} r^2 - 1\right)\Big],
%{\cal{L}}_{V} &=& \sum_{s=0}^{3}{{\cal{L}}_{V}^{(s)}},
\eqn
where   $\Omega_{ij} \equiv \delta^{\theta}_{i}\delta^{\theta}_{j}  + \sin^{2}\theta\delta^{\phi}_{i}\delta^{\phi}_{j}$
and  $A = A(r)$.  The expression for ${\cal{L}}_{V}$ is very complicated and shall not be given explicitly here.

 In the vacuum case, we have
 \bq
 \lb{4.3a}
 J^{t} = J_{A} = J_{\varphi} = 0,\;\;\; J_{i} = 0, \;\;\;
 \tau_{ij} = 0.
 \eq
 Then, the Hamiltonian and momentum constraints (\ref{eq1}) and (\ref{eq2}) reduce, respectively, to
 \bqn
 \lb{equ1}%4.3ba}
&&  \int{r^{2}e^{\nu} \left({\cal{L}}_{K} + {\cal{L}}_{V}\right)dr} = 0,\\
 \lb{equ2}%4.3bb}
&&  a(r) h'' + b(r) h' + c(r) h = 0,
 \eqn
 where
 \bqn
 \lb{4.3bc}
 a(r) &= & (1-\lambda)r^2f^2,\nb\\
 b(r) & =& \frac{1}{2}(1-\lambda)r f\left(4f - rf'\right),\nb\\
 c(r) &= &  - \frac{1}{2}(1-\lambda)\left[r^2\left(ff'' - {f'}^{2}\right) + 4f^2\right] - rff', ~~~~~~~~
 \eqn
 with
 \bq
 \lb{4.3bcc}
 f(r) = e^{-2\nu},\;\;\;
 h(r) = e^{\mu - \nu}.
 \eq
Equation (\ref{eq4a}) reads, % become,
\bqn
 \lb{equ2a}%4.3ca}
&&(1-\lambda)(rh'''+4h'')+d(r) h' + e(r) h=0, ~~~~~
\eqn
where
\bqn
\lb{4.3caa}
d(r) &=&\frac{1}{4rf^{2}}\Big\{4f^2+3(1-\lambda)r^2(f')^2\nb\\
&& ~~~~~~~~~
+4f[\Lambda_gr^2-1-(1-\lambda)r^2f'']\Big\}, \nb\\
e(r) &=& \frac{1}{4rf^{3}}\Big\{3(\lambda-1)r^2(f')^3+ff'\big[2(1-r^2\Lambda_g)\nb\\
&&  ~~~~~~~~~
+ (1-\lambda)r(4f'+5rf'')\big]+2f^2\big[(2\lambda-1)f'\nb\\
&&  ~~~~~~~~~
+ 4r\Lambda_g-(1-\lambda)r(2f''+rf''')\big]\Big\}.
\eqn
Equation (\ref{eq4b}), on the other hand,  yields,
\bqn
   \lb{equ3}%4.3cb}
&&  (rf)'  - \left(1 - \Lambda_gr^2\right) = 0,
 \eqn
while the dynamical equations (\ref{eq3}) read
\bqn
\lb{equ4}%4.3d1}
&& \left(\frac{A}{f^{1/2}}\right)'   + \frac{G(r)}{2rf^{3/2}} = 0, ~~~~~~~\\
\lb{equ4a}%4.3d2}
&& 2r^2fA'' +  {r} \left(2f + rf'\right) A' + {r}  \left(f' + 2 \Lambda_g r\right) A  + H(r) = 0, \nb\\
\eqn
where $G(r)$ and $H(r)$ are defined in Eq.(\ref{B.1}).

It should be noted that not all of the above equations are independent. In fact, Eq.(\ref{equ2a}) can be obtained from Eqs.(\ref{equ2}) and (\ref{equ3}), while
Eq.(\ref{equ4a}) can be obtained from Eqs.(\ref{equ4}), (\ref{equ2}) and (\ref{equ3}). Therefore, {\em in the present case
 there are only three independent differential equations,
 (\ref{equ2}),  (\ref{equ3}), and Eqs.(\ref{equ4}), for the three unknowns, $(f,\; h,\; A)$} \footnote{Certainly, such obtained solutions must satisfy the  global constraint (\ref{equ1}).}.
In particular, from Eq.(\ref{equ3}), we find that the general solution for $f$ is given by
\bq
\lb{fsolution}
f(r) = %- \frac{1}{2}\ln\left(
1-\frac{2B}{r} - \frac{1}{3}\Lambda_{g}r^{2}, %\right),
\eq
where $B$ is an integration constant.

Note that  the momentum constraint (\ref{equ2}) is a linear second-order ordinary differential equation for $h(r)$, and in principle one can integrate it to
find $h(r)$ for the general solution $f(r)$ given above. Once $h(r)$ is found, one can integrate Eq.(\ref{equ4})  to obtain $A$,
\bq
\lb{4.5}
A(r) = f^{1/2}(r)\left(A_{0} - \frac{1}{2}\int{\frac{G(r)dr}{rf^{3/2}(r)}}\right),
\eq
where $A_{0}$ is an integration constant, and $G(r)$ is given by Eq.(\ref{B.1}). %with $h = 0$.

On the other hand, for the general solution (\ref{fsolution}),   the potential ${\cal{L}}_{V}$ defined by Eq.(\ref{2.5a}) is given by \cite{GLLSW},
\bq
\lb{B.0}
  \mathcal{L}_V =  2\Lambda + \frac{1}{36 r^9 \zeta ^4}\left(\alpha_0 + \alpha_1 r + \alpha_2 r^3 + \alpha_3 r^9\right),
\eq
where
\bqn
\lb{B.0a}
  \alpha_0 & =& - 216B^3 \left(g_6+30g_8\right),\nb\\
    \alpha_1 &=& 3240 B^2 g_8,\nb\\ %\\
     \alpha_2 &=& 216 B^{2}\Big[\left(2g_5 +2 g_6-5 g_8\right)\Lambda_g   +g_3 \zeta ^2\Big],\nb\\
%   \\
    \alpha_3 &=& 8\Lambda_{g}\Big[4\left(9g_4 + 3g_5 + g_6\right)\Lambda_g^2 + 6\zeta^2\left(3g_2 +g_3\right)\Lambda_g\nb\\
    && ~~~~~~~  - 9 \zeta^4\Big].
\eqn

All the solutions with $\lambda = 1$ were found in \cite{GSW,AP}, and their global structures were systematically studied in \cite{GLLSW}.
Therefore, in the rest of this section, we consider only the case where    $\lambda \not= 1$. The case $\lambda \not= 1$ was also studied in
\cite{AP2}, but only approximate solutions were found.

When $\lambda \not= 1$,  a particular solution of Eq.(\ref{equ2}) is $h (r) = 0$.
Then, from Eqs.(\ref{B.1}) and (\ref{fsolution}) we can see that $A(r)$ now is independent of $\lambda$, and the corresponding solutions will be
the same as those given in the  case  $\lambda = 1, h = 0,\; f\not=0$ \cite{GSW,AP}. Therefore, in the following, we consider only the case
 where $h(r) \not= 0$. It is found convenient to  consider the four cases,  $\Lambda_{g} = 0 = B;\;    \Lambda_{g} = 0,\; B \not= 0;\;  \Lambda_{g} \not= 0, \; B = 0$;
 and $B \Lambda_g \not=0$,  separately.

 \subsubsection{$\Lambda_{g} = 0 = B $}

 In this case,  the momentum constraint (\ref{equ2}) reduces to
 \bq
 \lb{4.6a}
 r^2h'' + 2r h' - 2h = 0,
 \eq
 which has the general solution,
 \bq
 \lb{4.6b}
 h(r) = C_1 r + \frac{C_2}{r^2},
 \eq
 where $C_1$ and $C_2$ are two integration constants. Inserting it into Eq.(\ref{4.5}), we find that
 \bq
 \lb{4.6c}
 A(r) =A_0-\frac{3C_2^2}{8r^4}+\frac{1}{8}\left[3(1-3\lambda)C_1^2+2\Lambda\right] r^2.
 \eq

\subsubsection{$\Lambda_{g} = 0,\; B \not= 0 $}

When $\Lambda_{g} = 0$ and  $B \not= 0$, the momentum constraint (\ref{equ2}) reduces to
 \bqn
 \label{4.7}
&&  r\left(r - 2B\right) h'' + \left(2r - 5B\right)h' \nb\\
 && ~~~~  - \frac{2}{r}\left(\frac{r^{2} - 5Br + 5B^{2}}{r - 2B}  - B\varpi\right)h = 0, ~~~
 \eqn
 where $\varpi\equiv {1}/{(\lambda-1)}$.
Note that when $\lambda  = 1$, we must have $B = 0$, and
Eq.(\ref{4.7})  is identically satisfied for any $h(r)$, as noticed
previously.
When $\lambda \not= 1$, setting
$x = r/(2B),\; h(x)=h_0(r)h_1(x)$, Eq.(\ref{4.7}) takes the form,
  \bq
  \label{4.7b}
 x(1-x)\frac{d^2h_1}{dx^2}+p(x) \frac{dh_1}{dx}
 - q(x)h_1=0,
 \eq
 where
 \bqn
 \lb{4.7ba}
 p(x) &=& 2x(1-x)\frac{h_0^\prime}{h_0}+\frac{5-4x}{2},\nb\\
 q(x) &=& x(x-1)\frac{h_0^{\prime\prime}}{h_0}+\frac{4x-5}{2}\frac{h_0^\prime}{h_0}\nb\\
 && +\frac{4x^2-10x+5}{2x(1-x)}+\frac{\varpi}{x}.
 \eqn
 Assuming that Eq.(\ref{4.7b}) takes the form of the  hypergeometric differential equation,
  \bqn \label{4.7c}
x(1-x)\frac{d^2h_1}{dx^2}+\big[c-(a+b+1)x\big]\frac{dh_1}{dx}-abh_1=0,\nb\\
 \eqn
where $a, \; b $ and $c$ are constants, from Eq.(\ref{4.7ba})  we find
  \bqn \label{4.7da}
&& 2x(1-x)\frac{h_0^\prime}{h_0}+\frac{5-4x}{2}=c-(a+b+1)x,\\
  \label{4.7db}
&& x(x-1)\frac{h_0^{\prime\prime}}{h_0}+\frac{4x-5}{2}\frac{h_0^\prime}{h_0}+\frac{4x^2-10x+5}{2x(1-x)}+\frac{\varpi}{x}=-ab.\nb\\
 \eqn
Eq.(\ref{4.7da})  has the solution,
   \bqn \label{4.7e}
h_0(x)=(x-1)^{\frac{1}{4}(3+2a+2b-2c)}x^{\frac{1}{4}(2c-5)},
 \eqn
 for which  Eq.(\ref{4.7db}) is satisfied identically, provided that
   \bqn \label{4.7g}
(a- b)^2-9&=& 0, \\
2ab -c(a+b+1) + 2(5+\varpi) &=& 0,\\
4c^2-8c-45-16\varpi&=& 0.
 \eqn
A solution of the above equations is given by
\bqn
\lb{4.7h}
 a &=& \frac{1}{4}(7+\lambda_0), \;\;\;
 b= \frac{1}{4}(\lambda_0-5),\nb\\
 c&=& \frac{1}{2} (2+\lambda_0),\;\;\;
 \lambda_0=\sqrt{49+16\varpi}.
 \eqn
Thus,   the general solution of $h(r)$ takes the form,
  \bqn \label{4.8}
h(r)&=& h_0(r)\Bigg[a_1  \; F\left(a,b; c; x\right) \nb\\
&& + a_2\; x^{-\frac{1}{2}\lambda_0} %\nb\\ && \times
F\left(\frac{7-\lambda_0}{4},\frac{-5-\lambda_0 }{4};\frac{2- \lambda_0}{2}; x\right)\Bigg], \nb\\
 \eqn
where $a_1$ and $a_2$
are two integration (possibly complex) constants, and
${F}(a,b; c; z)$ is the hypergeometric function \cite{AS72}
 with ${F}(a,b;c; 0) = 1$. Inserting Eq.(\ref{4.7h}) into Eq.(\ref{4.7e}), we find that
\bqn
\lb{4.9}
 h_0(r)&=& \left(\frac{r}{2B}-1\right)^{1/2} \; \left(\frac{r}{2B}\right)^{\frac{\lambda_0 - 3}{4}}. \eqn

On the other hand, for $\Lambda_{g} = 0$   Eq.(\ref{4.5}) becomes,
 \bqn \label{4.10}
A(r)=\sqrt{1-\frac{2B}{r}}\left(A_0- \int{\frac{P(r)}{\sqrt{1-\frac{2B}{r}}}dr}\right),
 \eqn
where
 \bqn \label{4.11}
P(r)&=&\frac{1}{4(2B-r)^3}\Bigg\{\left[(9-25\lambda)B^2+2(1-2\lambda)r^2\right.\nb\\
&&+4Br(5\lambda-2)\big]h^2-2r(2B-r)\left[(5\lambda-1)B\right.\nb\\
&&-2r\lambda\big]hh^\prime-(r-2B)^2\left[\frac{4B}{r}+\frac{12B^3}{r^7\zeta^4}(22g_5\right.\nb\\
&& + 25g_6-20g_8)-\frac{2B^2}{r^6\zeta^4}(72g_5+81g_6-63g_8)\nb\\
&&\left.  -g_3\frac{2B^2}{r^4\zeta^2}-2\Lambda
r^2-r^2(1-\lambda)(h^\prime)^2\right]\Bigg\}.\nb\\
 \eqn

The Hamiltonian constraint (\ref{equ1}) now reads,
\bq
\lb{4.12}
\int{\frac{r^{2}\left({\cal{L}}_{V} + {\cal{L}}_{K}\right)dr}{\sqrt{1 - \frac{2B}{r}}}} = 0,
\eq
where ${\cal{L}}_{V}$ is given by Eq.(\ref{B.0}) with $\Lambda_g = 0$, and
% where
 \bqn
 \lb{4.13a}
 {\cal{L}}_{K} &=& \left(\frac{2}{r^2}+\frac{B^2}{(r-2B)^2r^2}\right)h^2+\frac{2Bhh^\prime}{2Br-r^2}\nb\\
 &&+(h^\prime)^2-\lambda\left(h^\prime+\frac{5B-2r}{r(2B-r)}h\right)^2.
 \eqn

\subsubsection{$B = 0,\;\;\; \Lambda_{g} \not= 0$}

When $B = 0,\; \Lambda_g \not= 0$, the momentum constraint (\ref{equ2}) reduces to
 \bqn
 \label{4.13}
&& h'' + \frac{1}{r} \left(1-\frac{3}{\Lambda_g r^2 -3}\right)h' \nb\\
 && ~~~ - \frac{18(\lambda-1)-3(5\lambda-7)\Lambda_g r^2 +(\lambda-3)\Lambda_g^2 r^4}{(\lambda-1) r^2(\Lambda_g r^2 -3)^2}h =
 0.\nb\\
 \eqn
 Note that when $\lambda = 1$ we have $\Lambda_g h = 0$.
 For $\lambda \not= 1$,  Eq.(\ref{4.13}) has the general solution,
 \bqn
 \label{4.14}
h&=&(1-z)\bigg\{\frac{b_1}{r^2}\; F\left(-\frac{\lambda_1}{2},\frac{\lambda_1}{2};-\frac{1}{2};z\right) \nb\\
 && +b_2\; rF\left(\frac{3-\lambda_1}{2},\frac{3+\lambda_1}{2};\frac{5}{2};z\right)\bigg\},\nb\\
 \eqn
 where  $b_1$ and $b_2$ are  constants, and
 \bq
 \lb{4.15}
 z=  \frac{1}{3} \Lambda_g r^2, \;\;\; \lambda_1=\sqrt{\frac{\lambda-3}{\lambda-1}}.
 \eq
 Then,  Eq.(\ref{4.5}) becomes,
 \bqn
 \lb{4.16}
 A(r)=\sqrt{3-\Lambda_gr^2}\left(A_0-
\int{\frac{Q(r)}{\sqrt{3-\Lambda_gr^2}}dr}\right),
  \eqn
where
 \bqn
 \label{4.17}
Q(r)&=&\frac{1}{12(3-\Lambda_gr^2)^3r}\Bigg\{9\bigg[\lambda(\Lambda_gr^2-6)^2-3\Big[6\nb\\
&&+(\Lambda_gr^2-4)\Lambda_gr^2\Big]\bigg]h^2+18(3-\Lambda_gr^2)r\Big[6\lambda\nb\\
&&-(\lambda+1)\Lambda_gr^2\Big]hh'+(3-\Lambda_gr^2)^2r^2\Bigg[6\Lambda_g\nb\\
&&-18\Lambda+9(\lambda-1)(h')^2+\frac{4\Lambda_g^2}{\zeta^2}(3g_2+g_3)\nb\\
&&+\frac{8\Lambda_g^3}{\zeta^4}(9g_4+3g_5+g_6)\Bigg]\Bigg\}.\nb\\
 \eqn
 The Hamiltonian constraint (\ref{equ1}), on the other hand,  takes the form,
 \bq
 \lb{4.17a}
 \int{\frac{r^{2}\left({\cal{L}}_{V} + {\cal{L}}_{K}\right)dr}{\sqrt{1 - \frac{1}{3}\Lambda_g r^2}}} = 0,
 \eq
 where ${\cal{L}}_{K}$ takes the same form of Eq.(\ref{4.13a}) but now with $h(r)$ given by Eq.(\ref{4.14}) and
 ${\cal{L}}_{V}$  given by,
  \bq
 \lb{4.18aa}
{\cal{L}}_{V} = 2\Lambda + \frac{\alpha_{3}}{36\zeta^{4}},
 \eq
 as   can be seen from Eqs.(\ref{B.0}) and (\ref{B.0a}).

 \subsubsection{$B \not= 0,\; \Lambda_g \not= 0$}

In this case, Eqs.(\ref{equ2}) becomes,
 \bqn
 \label{4.18}
&& r\left(6B-3r+\Lambda_g r^3\right) h'' + \left(15B-6r+\Lambda_g r^3\right)h' \nb\\
&& ~~~~~~~~~  -\left(\frac{6B-2\Lambda_gr^3}{r(\lambda-1)}+\frac{18(5B^2-5Br+r^2)}{r(6B-3r+\Lambda_g
 r^3)}\right.\nb\\
&&  ~~~~~~~~~   \left. +\frac{3\Lambda_g(16B-5r)+\Lambda_g^2r^6}{r(6B-3r+\Lambda_g
r^3)}\right)h =  0.
 \eqn
Setting \bq \lb{4.18a} r=2Bx, \;\;\; \Lambda_0 =
\frac{4}{3}B^2\Lambda_g, \eq
 and
$h(r)=h_0(x)h_1(x)$, where
 \bq
 \label{4.19}
h_1(x)=\exp{\int\frac{2c-5-2x\left[a+b-1+(\Lambda_0-e)x^2\right]}{4x(1-x+\Lambda_0)}dx},
\eq
we find that  Eq.(\ref{4.18}) reduces to
  \bqn
 \label{4.19b}
&&x(1-x+\Lambda_0x^3)h_0'' +\left[c-(a+b+1)x+ex^3\right]h_0'\nb\\
 && ~~~~~~~~~~~~~~~~~~~~~~~~~ -(ab+kx^2)h_0=0,
  \eqn
 but now with
 \bqn
 \lb{4.19c}
 a &=& \frac{\lambda_0+7}{4},\;\;\;
  b=\frac{\lambda_0-5}{4}, \nb\\
  c &=& \frac{\lambda_0+2}{2},\;\;\;
e=\frac{\Lambda_0}{2}(\lambda_0+5),\nb\\
 k&=&-\frac{3\Lambda_0}{8}(\lambda_0+7+8\varpi).
\eqn

When $c$ is not an integral, expanding $h_0(x)$ in the form, \bq
\lb{4.19d} h_0=
A_1\sum\limits_{i=1}^{\infty}a_ix^i+A_2\sum\limits_{i=1}^{\infty}b_ix^{i+1-c},
\eq where $A_1$ and $A_2$ are two constants, we find that in terms
of the two arbitrary constants $a_0$ and $b_0$,  the coefficients
$a_i$ and $b_i \; (i \not=0)$ are given by,
 \bqn
 \label{4.20}
a_1&=&\frac{ab}{c}a_0,\nb\\
a_2&=&\frac{ab(a+b+ab+1)}{2c(c+1)}a_0,\nb\\
a_3&=&\bigg[\frac{k}{3(c+2)}\nb\\
&&+\frac{ab(a+b+ab+1)(2a+2b+ab+4)}{6c(c+1)(c+2)}\bigg]a_0,\nb\\
a_4&=&\bigg[\frac{(k-e)ab}{4c(c+3)}+\frac{(a+b+ab+1)(2a+2b+ab+4)}{24c(c+1)(c+2)(c+3)}\nb\\
&&\times(3a+3b+ab+9)ab+\frac{k(3a+3b+ab+9)}{12(c+2)(c+3)}\bigg]a_0,\nb\\
b_1&=&\frac{ab+(1-c)(a+b-c+1)}{2-c}b_0,\nb\\
b_2&=&\frac{ab+(1-c)(a+b-c+1)}{2(2-c)(3-c)}\nb\\
&&\times\left[ab+(2-c)(a+b-c+2)\right]b_0,
\eqn
and
\bqn
  \label{4.21a}
 a_j&=&\frac{ab+(j-1)(a+b+j-1)}{j(j-1+c)}a_{j-1}~~~~~\nb\\
 &&-\frac{(j-3)(j\Lambda_0-4\Lambda_0+e)-k}{j(j-1+c)}a_{j-3}, ~~~~ j\geq 5\nb\\
 \label{4.21b}
 b_j&=&-\frac{(j-2-c)[(j-3-c)\Lambda_0+e]-k}{j(j-3+c)}b_{j-3}~~~~~\nb\\
 &&+\frac{ab+(j-c)(a+b+j-c)}{j(j-3+c)}b_{j-1}, ~~~~j\geq 3\nb\\
 \eqn
Note that one can always set $A_1 = A_2 = 1$,  by redefining the two arbitrary constants $a_0$ and $b_0$.

On the other hand, when $c=1+m$ (where $m$ is an integral), we
can use the Frobenius method to solve  Eq.(\ref{4.19b}). Let us first write
$h_0(x)$ in the form,
\bqn \label{4.22} h_0= A_1\sum\limits_{i=0}^{\infty}a_ix^i+A_2
x^{-m}\left[\ln
x\sum\limits_{i=m}^{\infty}\bar{a}_ix^i+\sum\limits_{i=0}^{\infty}b_ix^i\right],\nb\\
\eqn
where
\bqn \label{4.23}
\frac{c-(a+b+1)x+ex^3}{1-x+\Lambda_0x^3}=\sum\limits_{i=0}^{\infty}c_ix^i,\nb\\
\frac{-abx-kx^3}{1-x+\Lambda_0x^3}=\sum\limits_{i=0}^{\infty}d_ix^i,
\eqn then, we can obtain the coefficients $a_i$, $\bar{a}_i$ and
$b_i$ in terms of the two arbitrary constants $c_0$ and $d_0$, which
are given by \bqn \label{4.24}
a_i&=&-\frac{\sum\limits_{k=1}^{i}a_{i-k}[c_k(i-k)+d_k]}{i(i-1+c_0)+d_0},\nb\\
\bar{a}_i&=&-\frac{\sum\limits_{k=1}^{i}\bar{a}_{i-k}[c_k(i-k-m)+d_k]}{(i-m)(i-m-1+c_0)+d_0},\nb\\
b_i&=&-\frac{\bar{a}_i(c_0-1-2m+2i)}{(i-m)(i-m-1+c_0)+d_0}\nb\\
&&-\frac{\sum\limits_{k=1}^{i}\{\bar{a}_{i-k}c_k+b_{i-k}[c_k(i-k-m)+d_k]\}}{(i-m)(i-m-1+c_0)+d_0}.\nb\\
\eqn

\section{Failure in Solar System Tests}
\renewcommand{\theequation}{4.\arabic{equation}} \setcounter{equation}{0}

 The solar system tests are usually written in terms of the Eddington parameters, by following the so-called
 ``parameterized post-Newtonian" (PPN) approach, introduced initially by Eddington \cite{Edd}. These parameters are often
 written in terms of the line element   in its diagonal form,
 \bq
 \lb{6.0a}
 ds^{2} = - e^{2\Psi(r)}d\tau^2 + e^{2\Phi(r)}dr^2 + r^2d\Omega^2.
 \eq
  Then, the gravitational
 field, produced by a  point-like and motion-less particle with mass $M$, is given by
 \bqn
 \lb{6.0b}
 e^{2\Psi} &=& 1 - 2\left(\frac{GM}{c^2 r}\right) + 2\big(\beta - \gamma\big) \left(\frac{GM}{c^2 r}\right)^2 + ...,\nb\\
 e^{2\Phi} &=& 1 + 2\gamma \left(\frac{GM}{c^2 r}\right) +  ..., 
 \eqn
where $\beta$ and $\gamma$ are the Eddington parameters. For the solar
 system,    we have $r_g \equiv GM_{\bigodot}/c^2 \simeq 1.5$ km, and its radius is
  $r_{\bigodot} \simeq 1.392\times 10^{6}$ km. So, within the solar system
    the dimensionless quantity $\chi [\equiv GM/(rc^2)]$   in most cases  is much less than one,  $\chi  \leq r_{g}/r_{\bigodot} \le
    10^{-6}$.
 The Shapiro delay of the   Cassini probe \cite{BT}, and the solar system ephemerides
\cite{WTB} yield, respectively, the bounds \cite{RJ},
\bqn
\lb{6.2}
\gamma - 1&=& (2.1 \pm 2.3)\times 10^{-5},\nb\\
\beta - 1&=& (-4.1  \pm 7.8)\times 10^{-5}.
\eqn
GR predicts $\beta = 1 = \gamma$ precisely.
To study the solar system tests in the HL theory, we may first transform the above experimental results in terms of  the ADM line element
with the projectability condition,
\bq
\lb{6.2a}
ds^{2} = - dt^2 + e^{2\Omega}\left(dr + e^{\Gamma - \Omega} dt\right)^2 + r^2d\Omega^2,
\eq
for which it can be shown that   \cite{GSW}, 
 \bqn
\lb{6.1}
\Gamma  &=& \frac{1}{2}\ln\left\{2c^{2}\Bigg[\left(\frac{GM}{c^{2}r}\right) - \big(\beta - \gamma\big)\left(\frac{GM}{c^{2}r}\right)^{2} + ...\Bigg]\right\},\nb\\
 \Omega &=& \big(\gamma - 1\big)  \left(\frac{GM}{c^{2}r}\right)  + ....
\eqn
 
In the case $\lambda =1$, two different identifications were prescribed. One was to consider $A$ as part of the metric  via the replacement \cite{HMT},
\bq
\lb{6.3a}
N  \rightarrow N - \frac{1}{c^2} A.
\eq
With such an identification, the diagonal solution \cite{HMT,AP,GSW},
\bqn
\lb{6.3b}
&& N = 1, \;\;\; N^i = 0,\;\;\; f = 1 - \frac{2m}{r},\nb\\
&& A = 1 - A_0\sqrt{1 - \frac{2m}{r}}, \;\;\; \varphi = 0,\; (\lambda = 1),
\eqn
produces exactly the Schwarzschild solution in the form (\ref{6.0a}) with $\Psi = -\Phi = \frac{1}{2}\ln(f)$. Note that in writing Eq.(\ref{6.3b}), the speed of light appearing in
Eq.(\ref{6.3a}) had been  set to one. As a result, the theory is consistent with observations \cite{HMT}. 

However, the solution (\ref{6.3b}) is not unique, and there exists a larger class of non-diagonal solutions given by \cite{GSW},
 \bqn
\lb{4.3f}
\Gamma &=& \ln{h(r)} \nb\\
&=&  \frac{1}{2}\ln\left(\frac{2B}{r} + \frac{1}{3}\Lambda r^2 - 2A(r) + \frac{2}{r}\int^{r}{A(r') dr'}\right),\nb\\
\Omega &=& 0,\;\;\;  \varphi = 0, \;  (\lambda = 1),
\eqn
where the gauge field $A(r)$ is undetermined.   If one does not consider the gauge field $A$ as a part of metric \cite{GSW}, but simply considers it as representing 
a degree of freedom  of the gravitational
field, as the Brans-Dicke scalar field  in  the Brans-Dicke theory \cite{BD}, one finds that the above solutions are consistent with all the solar system tests, provided that \cite{GSW}, 
\bq
\lb{6.3}
A(r) = {\cal{O}}\left[\left(\frac{GM}{c^{2}r}\right)^2\right].
\eq
 In the rest of this section we shall  follow the second prescription, i.e., setting directly,
\bq
\lb{idd}
(\Gamma, \Omega) = (\mu, \nu),
\eq
 and verify whether this prescription can be generalized to the case $\lambda \not= 1$. 
 As to be shown below, the answer is unfortunately negative. 
 
 To this goal, we first note that   the cosmological constant $\Lambda$
has negligible effects within the solar system. In addition, the spatial curvature of the solar system is negligible. In fact, for the metric (\ref{6.2a}), it takes the form,
\bqn
\lb{6.3d}
R &=& \frac{2}{r^2}\Big[1 - e^{-2\Omega}\left(1 - 2r\Omega'\right)\Big]\nb\\
&\simeq& \frac{8(\gamma -1)^2}{r_g^2}\left(\frac{GM}{c^2r}\right)^3, %\chi^3 < \frac{10^{-28}}{{\mbox{km}}^2}.
\eqn
for $r >> r_g \equiv GM/c^2$. Note that in writing the last step of the above equation, we had used Eq.(\ref{6.1}). Thus, in the solar system we have $\Lambda_g = R/2 < 10^{-28}\; {\mbox{km}}^{-2}$. 
Therefore, without loss of generality,
we  set 
\bq
\lb{6.3c}
\Lambda = \Lambda_g = g_{s} = 0,\; (s \ge 2).
\eq
Then, the solutions are those given by Eqs.(\ref{fsolution}) and (\ref{4.8}), from which we find that
\bqn
\lb{6.4}
\nu &=&- \frac{1}{2} \ln\left(1 - \frac{2B}{r}\right) \nb\\ 
&\simeq& \epsilon \left(\frac{GM}{c^{2}r}\right)
 +   \epsilon^2\left(\frac{GM}{c^{2}r}\right)^2  \nb\\
 && +  {\cal{O}}\left[\left(\frac{GM}{c^{2}r}\right)^3\right],
\eqn
where $ \epsilon \equiv  {Bc^2}/{(GM)}$. Comparing the above with Eq.(\ref{6.1}), we find that   
\bqn
\lb{6.5}
\gamma - 1 =  \epsilon = \frac{Bc^2}{GM} \le 10^{-4},
\eqn
for $M = M_{{\bigodot}}$. As a result, we have
\bq
\lb{6.6}
x = \frac{r}{2B} = \frac{1}{2(\gamma - 1)}  \left(\frac{GM}{c^{2}r}\right)^{-1} \gg 1.
\eq
 From the relations,
\bqn
\lb{6.7}
F(a, b; d; x) &=& (1-x)^{-b} F\left(b, d-a; d; \frac{x}{x-1}\right),\nb\\
F(a, b; d; 1) &=&    \frac{\Gamma(d)\Gamma(d-a-b)}{\Gamma(d-a)\Gamma(d-b)},
\eqn
where the last expression holds only for $d \not= 0, -1, -2, ..., {\mbox{Re}}(d - a - b) > 0$,  we find from Eq.(\ref{4.8}) that $h(r)$ has the asymptotical form,  
\bq
\lb{6.8}
h(r) \simeq D_{1} r,\; (x \gg 1), 
\eq
 with
\bqn
\lb{6.9}
D_{1} &=& \frac{1}{2B}\Big[a_1(-1)^{b}F(b, c-a; c; 1)\nb\\
&& ~~~~~+ a_2 (-1)^{\hat{b}}F(\hat{b}, \hat{c} - \hat{a}; \hat{c}; 1)\Big],\nb\\
\hat{a} &=& \frac{7-\lambda_0}{4},\;\;
\hat{b} = - \frac{5+\lambda_0}{4},\;\;
\hat{c} = \frac{2- \lambda_0}{2}.
\eqn
Note that Eq.(\ref{6.8}) can be also obtained directly from Eq.(\ref{4.7c}), which reads
\bq
\lb{6.10}
x^2 h_{1}'' + (a+b+1)x h_{1}' + ab h_{1} = 0,
\eq
for $x \gg 1$.  Eq.(\ref{6.10})  has the general solution,
\bq
\lb{6.11}
h_1(x) = d_1 x^{(5-\lambda_0)/4} + d_2 x^{-(7 + \lambda_0)/4}, %\; (x \gg 1),
\eq
where $d_1$ and $d_2$ are two integration constants. On the other hand, from Eq.(\ref{4.7e}) we find that
\bq
\lb{6.12}
h_{0}(x) \simeq x^{(\lambda_0-1)/4}. %\; (x \gg 1).
\eq
Then, we obtain
\bq
\lb{6.13}
h(x) = h_0(x) h_1(x)  \simeq  d_1 x + \frac{d_2}{x^2} \simeq d_1 x, %\; (x \gg 1)
\eq
 which is precisely the solution given by Eq.(\ref{6.8}) with $D_1 = d_1/(2B)$.  Hence, we obtain
\bqn
\lb{6.14}
\Gamma(r) &=& \frac{1}{2}\ln\left(\frac{h^2}{f}\right)\nb\\
&=& \frac{1}{2}\ln\Bigg\{\left(\frac{c^2 r}{GM}\right)^{2}\Bigg[1 + \left(\frac{2c^2B}{GM}\right) \left(\frac{GM}{c^2 r}\right) \nb\\
&& + \left(\frac{2c^2B}{GM} \right)^2 \left(\frac{GM}{c^2 r}\right)^2 + {\cal{O}}\left(\chi^{3}\right)\Bigg]\Bigg\} \nb\\
&& + \ln\left(\frac{D_1GM}{c^2}\right).
\eqn
This is quite different from that given by Eq.(\ref{6.1}) with any choice of $a_1,\; a_2,\; B$, as long as $\lambda$ is not exactly equal to one. 
Therefore,  
{ the static vacuum solutions given by Eqs.(\ref{fsolution}) and (\ref{4.8}) with the condition (\ref{6.3c}) %$\Lambda_g = \Lambda = 0$ and $\lambda \not= 1$
is inconsistent with the solar system tests, when the prescription (\ref{idd}) is used.}

\section{Most general ansatz with spherical symmetry}
\renewcommand{\theequation}{5.\arabic{equation}} \setcounter{equation}{0}

In the previous section, based on the stationary ansatz (\ref{4.1}), we
have seen that the prescription (\ref{idd}) leads to failure in solar
system tests for $\lambda\ne 1$, however small $|\lambda-1|$ is. Hence,
in the next section we shall consider another prescription. Before that,
however, in this section let us consider the most general ansatz with
spherical symmetry and show that the prescription (\ref{idd}) never
recovers the Schwarzschild geometry in the $\lambda\to 1$ limit with
$\Lambda_g=0$. This confirms that a prescription beyond (\ref{idd}) is
absolutely necessary.

In order to find the most general ansatz, note that one can always
choose time and spatial coordinates so that $N=1$ and $N^i=0$ at least
locally.  One can also set $\varphi=0$ by the $U(1)$ gauge freedom. With  
\begin{equation}
\lb{metricB}
 N = 1, \quad N^i = 0, \quad \varphi=0,
\end{equation}
it is obvious that the most general ansatz with spherical symmetry is \footnote{
One must not confuse with the function $B(t,x)$ used in this section and the constant $B$ used in the expression of $f(r)$ in the previous and next
 sections.}, 
\begin{equation}
\lb{metricC}
 g_{ij}dx^idx^j = e^{2B(t,x)}dx^2 + e^{2C(t,x)}d\Omega^2, 
  \quad  A = A(t,x). 
\end{equation}
Independent equations are the equation of motion for the gauge field
$A$, the $x$-component of the momentum constraint and the $xx$-component
of the dynamical equation.

The equation of motion for the gauge field $A$ is written as
\begin{equation}
 \partial_x\left[e^{-2B+3C}(\partial_xC)^2
	    +\frac{1}{3}\Lambda_g e^{3C} - e^{C}\right] = 0,
\end{equation}
leading to the general solution
\begin{equation}
 B = \frac{3}{2}C + \frac{1}{2}\ln
  \left[\frac{(\partial_xC)^2}{F(t)+e^C-(\Lambda_g/3)e^{3C}}\right],
  \label{eqn:Bsol-dynamical}
\end{equation}
where $F(t)$ is an arbitrary function of time. The momentum constraint
is
\bqn
&& 
 \partial_x\partial_tC +
  \partial_xC\partial_tC-\partial_xC\partial_tB\nb\\
  && ~~~~~~~~
  + (\lambda-1)
  \left[\partial_x\partial_tC+\frac{1}{2}\partial_x\partial_tB\right] = 0.
\eqn
By using the solution (\ref{eqn:Bsol-dynamical}), this equation is
reduced to an equation for $C$:
\begin{eqnarray}
 &  &\partial_t\left[\Lambda_g e^{2C}-3Fe^{-C}\right]
  \nonumber\\
& & = (\lambda-1)
 \left\{c_1
  \left[\frac{\partial_x^2\partial_tC}{(\partial_xC)^2}
   - \frac{(\partial_x\partial_tC)(\partial_x^2C)}
   {(\partial_xC)^3}\right] \right.
  \nonumber\\
& & \left.
   + c_2\frac{\partial_x\partial_tC}{\partial_xC}
   + c_3\partial_tC + c_4\right\},
  \label{eqn:ode-for-C-dynamical}
\end{eqnarray}
where
\begin{eqnarray}
 c_1 & = & \frac{1}{\Delta}\Bigg[3F^2e^{-C}+2(3-\Lambda_g e^{2C})F\nb\\
 && ~~~~~~~
  +\frac{1}{3}e^{-C}(\Lambda_g e^{3C}-3e^C)^2\Bigg],\nb\\
%  {F+e^C-(\Lambda_g/3)e^{3C}},\nonumber\\
 c_2 & = & \frac{1}{6\Delta}\Big[63F^2e^{-C}+3(39-11\Lambda_g e^{2C})F\nb\\
 && ~~~~~~~
  +2(2\Lambda_g^2 e^{5C}-15\Lambda_g e^{3C}+27e^{C})\Big],\nb\\
  %{F+e^C-(\Lambda_g/3)e^{3C}},\nonumber\\
 c_3 & = & \frac{1}{2\Delta}
  \Big[3(3\Lambda_g e^{2C}-1)F +4\Lambda_g e^{3C}\Big], \nb\\
 % {F+e^C-(\Lambda_g/3)e^{3C}},\nonumber\\
 c_4 & = & \frac{3}{2\Delta}{(1-\Lambda_g e^{2C})\partial_t F}, \nb\\
  \Delta &=& F+e^C-(\Lambda_g/3)e^{3C}.
\end{eqnarray}
Finally, the dynamical equation can be considered as an equation
determining the gauge field $A$. With the prescription (\ref{idd}) where
$A$ does not participate in the geometry nor in solar system tests, the
dynamical equation is not of our interest.

Now let us expand $C$ by $(\lambda-1)$ as 
\begin{equation}
 C(t,x) = \sum_{n=0}^{\infty}C_n(t,x)(\lambda-1)^n. 
  \label{eqn:expansion-C}
\end{equation}
We shall see below that the momentum constraint equation
(\ref{eqn:ode-for-C-dynamical}) can be solved iteratively order by order
in the $(\lambda-1)$ expansion, under the condition
(\ref{eqn:condition-for-expansion}) below.

First, the zeroth order solution $C_0$ is obtained as a solution to the 
following algebraic equation 
\begin{equation}
 \Lambda_g e^{2C_0(t,x)}-3F(t)e^{-C_0(t,x)} = G(x),
  \label{eqn:Csol-zeroth}
\end{equation}
where $G(x)$ is an arbitrary function of $x$. Next, let us show by
induction that the $n$-th order solution $C_n(t,x)$ can be obtained by
solving (\ref{eqn:ode-for-C-dynamical}) order by order. For this
purpose, let us expand the expression inside the squared bracket on the
left hand side of (\ref{eqn:ode-for-C-dynamical}) by $(\lambda-1)$ as
\begin{equation}
  \Lambda_g e^{2C}-3F(t)e^{-C} = 
   \sum_{n=0}^{\infty}{\cal G}_n(t,x)(\lambda-1)^n,
\end{equation}
according to the expansion (\ref{eqn:expansion-C}). It is easy to
understand that ${\cal G}_n$ has the form 
\begin{equation}
 {\cal G}_n = \left(2\Lambda_g e^{2C_0}+3F(t)e^{-C_0}\right)C_n
  + \tilde{\cal G}_n, \label{eqn:expand-calGn}
\end{equation}
where $\tilde{\cal G}_n$ depends only on $C_i$
($i=1,2,\cdots,n-1$). Thus, provided that 
\bqn
&& 2\Lambda_g e^{2C_0}+3F(t)e^{-C_0} \ne 0, \nb\\
&& F+e^{C_0}-(\Lambda_g/3)e^{3C_0} \ne 0, \nb\\
&&  \partial_xC_0\ne 0,  
\label{eqn:condition-for-expansion}
\eqn
we obtain
\begin{eqnarray}
 C_n(t,x) & = & - 
  \frac{1}{\Delta_1}
  \left[\tilde{\cal G}_n[C_1,\cdots,C_{n-1};t,x)
   \right.\nonumber\\
 & & \left.
   - \int_{t_0}^tdt' S_n[C_1,\cdots,C_{n-1};t',x)
       \right], \nb\\
\Delta_1 &=& 2\Lambda_g e^{2C_0}+3F(t)e^{-C_0}, 
 \label{eqn:Cnsol}
\end{eqnarray}
where $S_n$ is the $n$-th order part of the right hand side of
(\ref{eqn:ode-for-C-dynamical}), which also depends only on $C_i$
($i=1,2,\cdots,n-1$), and $t_0$ is an initial time. Note that the
initial value of $C_n$ at $t=t_0$ has been set to zero by redefinition
of $G(x)$ and that the change due to the shift of the initial time $t_0$
corresponds to redefinition of $G(x)$. From this result, it is obvious
by induction that the solution of the form (\ref{eqn:expansion-C}) can
be obtained up to any order of the expansion.

Let us consider the zero-th order solution (\ref{eqn:Csol-zeroth}) with
$\Lambda_g=0$. In order for the expansion w.r.t. ($\lambda-1$) to make
sense, the condition (\ref{eqn:condition-for-expansion}) must be
satisfied. In particular, $F(t)$ should be non-vanishing. Otherwise, the
denominator on the r.h.s. of (\ref{eqn:Cnsol}) would vanish. We thus
assume that $F(t)\ne 0$.

We would like to see if the Schwarzschild geometry is recovered in the
limit $\lambda\to 1$ with $\Lambda_g=0$ or not. One of the simplest ways
is to calculate the $4$-dimensional Einstein tensor for the
$4$-dimensional metric 
\begin{equation}
 ds_4^2 = -dt^2 + e^{2B_0(t,x)}dx^2 + e^{2C_0(t,x)}d\Omega^2, 
\end{equation}
where 
\begin{equation}
 B_0 = \frac{1}{2}\ln
  \left[\frac{27F(t)^2(\partial_xG(x))^2}{G(x)^4(3-G(x))}\right],
  \quad
  C_0 = \ln\left[\frac{-3F(t)}{G(x)}\right]. 
\end{equation}
Non-vanishing components of the $4$-dimensional Einstein tensor are
\begin{eqnarray}
 G^{(4)t}_{\ t} & = & -\frac{3(\partial_tF)^2}{F^2}, \nonumber\\
 G^{(4)x}_{\ x} & = & -\frac{1}{F^2}
  \left[(\partial_tF)^2+2F\partial_t^2F+\frac{G^3}{27}\right], \nonumber\\
 G^{(4)\theta}_{\ \theta} & = & -\frac{1}{F^2}
  \left[(\partial_tF)^2+2F\partial_t^2F-\frac{G^3}{54}\right]. 
\end{eqnarray}
In order to recover the Schwarzschild metric, all of these components must
vanish, leading to $\partial_tF=G=0$. However, in this case the
regularity of $C_0$ implies that $F=0$, contradicting with the
assumption $F\ne 0$. Note that $F\ne 0$ is a necessary condition for the
continuity of the $\lambda\to 1$ limit.

If we set $F=0$ then the $\lambda\to 1$ limit is singular. Thus, the
zero-th order solution is not obtained as the $\lambda\to 1$ limit of
a solution with $\lambda\ne 1$. Instead, it represents a solution with
exactly $\lambda=1$. In this case, (\ref{eqn:Csol-zeroth}) with
$\Lambda_g=0$ implies that $G=0$ and leaves $C_0$ unspecified. There is 
a choice of $C_0$ giving rise to the Schwarzschild metric.

Just for completeness, let us consider the case with $F=0$ and
$\Lambda_g\ne 0$. In this case the $\lambda\to 1$ limit is
continuous. However, the zeroth order solution is 
\begin{equation}
 B_0 = \frac{1}{2}\ln\frac{3(\partial_xG)^2}{4\Lambda_g G(G-3)},
  \quad
  C_0 = \frac{1}{2}\ln\frac{G}{\Lambda_g},
\end{equation}
resulting in 
\begin{equation}
 G^{(4)t}_{\ t} = -\Lambda_g, \quad
 G^{(4)x}_{\ x} = 
 G^{(4)\theta}_{\ \theta} = -\frac{1}{3}\Lambda_g. 
\end{equation}
Hence, the Schwarzschild metric is not recovered in this case unless the 
limit $\Lambda_g\to 0$ is taken. If we take this limit then the
$\lambda\to 1$ limit becomes singular. Thus, again, the Schwarzschild
solution is not obtained as the $\lambda\to 1$ limit of a solution with
$\lambda\ne 1$. Instead, it represents a solution with exactly
$\lambda=1$.

In summary, if $\lambda=1$ and $\Lambda_g=0$ exactly then the
Schwarzschild metric is one of solutions. However, if we consider
$\lambda\ne 1$ and $\Lambda_g=0$, then the Schwarzschild metric is never
recovered in the limit $\lambda\to 1$. This conclusion is based on the
prescription (\ref{idd}) and thus implies that a prescription beyond
(\ref{idd}) is absolutely necessary.

\section{Physical Interpretation of $A$ and $\varphi$}
\renewcommand{\theequation}{6.\arabic{equation}} \setcounter{equation}{0}

In  this section, we shall show that a proper generalization of the prescription of (\ref{6.3a}) can lead to  solutions that are consistent with solar system tests
even for $\lambda \not= 1$. From such a generalization, the physical and geometrical interpretations of the gauge field $A$ and Newtonian prepotential $\varphi$
also become clear. 

\subsection {General Coupling of $A$ and $\varphi$ to Metric}

To the above claim, we first note that under the U(1) transformations, the ADM quantities transform as \cite{HMT,HW},
\bqn
\lb{Ugauge}
\delta_{\alpha}{N} &=& 0,\;\;\;
\delta_{\alpha}N_{i}  = N\nabla_{i}\alpha,\;\;\;
\delta_{\alpha}g_{ij} = 0,\nb\\
\delta_{\alpha}A &=&\dot{\alpha} - N^{i}\nabla_{i}\alpha,\;\;\;
\delta_{\alpha}\varphi = - \alpha,
\eqn
where $\delta_{\alpha}F = \tilde{F} - F$, $\alpha [= \alpha(t, x)]$ is   the generator  of the local $U(1)$ gauge symmetry. From the above we find that
\bqn
\lb{7.1}
\delta_{\alpha}{\cal{A}} &=& \dot{\alpha} - N^{i}\nabla_{i}\alpha, \nb\\
 \delta_{\alpha}\sigma &=& 0, \;\;\; \delta_{\alpha}{\cal{N}}^i = 0,
\eqn
where ${\cal{A}}$ is defined in Eq.(\ref{ident}), and 
\bq
\lb{eq0}
\sigma \equiv A - {\cal{A}},\;\;\;
{\cal{N}}^{i} \equiv N^i + N\nabla^i\varphi.
\eq

If we require that the line element $ds^2$ be gauge-invariant not only under ${\mbox{Diff}}(M, {\cal{F}})$
(\ref{1.2}),  but also under the enlarged symmetry (\ref{symmetry}),   then $ds^2$ defined by,
\bq
\lb{id}
ds^2 \equiv - {\cal{N}}^2c^2dt^2 + g_{ij}\left(dx^i + {\cal{N}}^idt\right)\left(dx^j + {\cal{N}}^j dt\right),
\eq
has the desired properties, where
\bq
\lb{id.a}
{\cal{N}} \equiv N - \frac{\upsilon}{c^2}\left(A - {\cal{A}}\right),
\eq
where $\upsilon$ is a dimensionless coupling constant subjected possibly to radiative corrections. Similar to $N$, such defined ${\cal{N}}$ is also dimensionless,
 $[{\cal{N}}] = 0$. With this prescription, one can see that the Newtonian prepotential $\varphi$ is 
tightly related to the shift vector ${\cal{N}}^i$, while the geometrical lapse function ${\cal{N}}$ is related to both $A$ and $\varphi$. 
 In addition, since
\bqn
\lb{id.b}
\left[dx\right] &=& -1, \;\;\; \left[dt\right] = - z,\;\;\; \left[c\right] = \frac{\left[dx\right]}{\left[dt\right]} = z-1,\nb\\
\left[N\right] &=& 0,\;\;\;  \left[N^i\right] = z-1,\;\;\; \left[g_{ij}\right] = 0,\nb\\
\left[A\right] &=& \left[{\cal{A}}\right] = 2(z-1), \;\;\; \left[\varphi\right] = z -2,\nb\\
\left[\alpha\right] &=& z-2,
\eqn
we find that
\bq
\lb{id.c}
\left[ds\right] = -1,
\eq
i.e., it has the dimension of length. Moreover, with the gauge choice $\varphi = 0$ and setting $\upsilon = 1$, Eq.(\ref{id.a}) reduces to Eq.(\ref{6.3a}).

In the Newtonian limit, we have \cite{DInverno,HMT}
\bq
\lb{id.d}
g_{00} = - \left( 1 + \frac{2\phi}{c^2} + {\cal{O}}(\epsilon) \right),\;\;\;
g_{0i} = {\cal{O}}(\epsilon),
\eq
in the coordinates $x^\mu = (ct, x^i)$, where $\epsilon \equiv |v/c| \ll 1$, and $v$ denotes the typical velocity of the system concerned.  Comparing it with the metric
given by Eqs.(\ref{eq0})-(\ref{id.a}), we find that the Newtonian potential $\phi$ is given by
\bqn
\lb{id.e}
&& \phi =  - \upsilon(A +\dot{\varphi}) - \frac{1}{2}N^iN_i \nb\\ 
&& ~~~~~+ \left(\upsilon - 1\right)\left(N^i + \frac{1}{2}\nabla^i\varphi\right)\nabla_i\varphi,
\eqn
with
\bq
\lb{id.f}
 N = 1,\;\;\; \frac{1}{c}\left|2N_i + \nabla_i\varphi\right| =  {\cal{O}}(\epsilon).
\eq

To study further   the meaning of the above prescription and the physical interpretations of the gauge field $A$ and Newtonian prepotential $\varphi$,
 let us turn to the solar system tests again. 

\subsection{Solutions with the Gauge $A = 0$}

For the ADM decomposition (\ref{4.1}) without fixing the U(1) gauge,
there are  three independent equations,  given by Eqs.(\ref{D.1}) - (\ref{D.3}).   To solve these equations, let us   first note that the prescriptions  of Eqs.(\ref{eq0}) - (\ref{id.a}) do not change
 the spatial metric  $g_{ij}$. As a result, the constraint on the spatial curvature $R$ takes the same form of Eq.(\ref{6.3d}). Therefore, in the present case the condition (\ref{6.3c}) can be 
 still imposed safely.  In particular, with the gauge $A = 0$  [cf. Appendix C for different gauge choices.],
Eqs.(\ref{D.1}) and   (\ref{D.2}) for $\lambda = 1$ have the solutions,
\bq
\lb{7.2}
f = 1 - \frac{2B}{r},\;\;\; h = - f \varphi',\; (\lambda = 1,\; A = 0),
\eq
where $\varphi$ must satisfy the dynamical equation  Eqs.(\ref{D.3}), which now reads,
\bqn \lb{7.3}
\left(1-\frac{2B}{r}\right)^2\left[(\varphi')^2\right]'+\left(1-\frac{2B}{r}\right)\frac{B}{r^2}(\varphi')^2+\frac{2B}{r^2}=0.\nb\\
\eqn
The general solutions are given by,
\bqn\lb{7.4}
\varphi(r)=\varphi_0\pm\int{\left(\frac{2r}{r-2B}+\varphi_1\sqrt{\frac{r}{r-2B}}\right)^{1/2} dr}, ~~~
\eqn
where $\varphi_0$ and $\varphi_1$ are integrations constants. Substituting the above into Eq.(\ref{eq0}) we find that
${\cal{N}}^i = 0$. Then,  Eq.(\ref{id}) reduces to,  
\bq \lb{7.5}
ds^2 = - {\cal{N}}^2 dt^2 + \frac{dr^2}{f(r)} + r^2
d^2\Omega,
\eq
where
\bqn
 \lb{7.6}
{\cal{N}}^2 &=& \frac{1}{4}\left[2- \upsilon\left(2 -\frac{\varphi_1(\epsilon\chi-1)}{\sqrt{1-\epsilon\chi}}\right)\right]^2,\nb\\
f &=& 1 - \frac{2B}{r}, \; (\lambda = 1, \; A = 0).
\eqn
When $\chi \ll 1$, we have
\bqn \lb{7.7}
{\cal{N}}^2  
&=& \Upsilon^2\Bigg(1 -\frac{C_1\upsilon\epsilon}{2\Upsilon}\chi -\frac{C_1(\upsilon-1)\upsilon\epsilon^2}{8\Upsilon^2}\chi^2 + {\cal{O}}\left(\chi^3\right)\Bigg),\nb\\
\frac{1}{f}  &=& 1 + \epsilon \chi + \epsilon^2\chi^2 + {\cal{O}}\left(\chi^3\right),
\eqn
where    $\Upsilon \equiv \upsilon(1 + \varphi_1/2) -1$. The factor
$\Upsilon$ appearing in the expression of 
${\cal{N}}$ can be dropped by rescaling $ t \rightarrow \Upsilon t$. Then,
comparing Eq.(\ref{7.7}) with Eq.(\ref{6.0b}) we find that
\bqn \lb{7.8}
B&=& \gamma \left(\frac{GM}{c^2}\right),\;\;\;
\beta =\frac{1}{2}(\gamma+1),\nb\\
\upsilon &=& \frac{2}{2 + (\gamma-1)\varphi_1}.
\eqn
For $\varphi_1 \simeq {\cal{O}}(1)$, we obtain the constraint 
$|\upsilon-1|< {\cal{O}}\left(10^{-5}\right)$
from (\ref{6.2}). For extremely large value of $\varphi_1$, say 
$\varphi_1 \simeq {\cal{O}}\left(10^5\right)$, 
$|\upsilon-1| \simeq {\cal{O}}\left(1\right)$ is also allowed. However,
we consider this large value of $\varphi_1$ unrealistic and consider the
case with $\varphi_1= {\cal{O}}\left(1\right)$ only. Note that the Schwarzschild 
solution corresponds to $B ={GM}/{c^2},\; \upsilon = 1$.

\subsection{Solutions with the Gauge $\varphi \cdot A \not = 0$}

On the other hand, in the case $\lambda \not= 1$ let us consider the gauge $h = 0$. Then, %Eq.(\ref{D.2}) is satisfied identically,
 Eqs.(\ref{D.1}) and (\ref{D.2})  yield, 
\bqn \label{7.9}
f(r) &=& 1 - \frac{2B}{r},\nb\\
\varphi(r)&=& \int r^{\frac{1-\lambda_0}{4}}\Bigg[b_1  \; F\left(\frac{9-\lambda_0}{4};\frac{-3-\lambda_0}{4}; \frac{2-\lambda_0}{2}; x\right) \nb\\
&& + b_2\;r^{\frac{\lambda_0}{2}}
F\left(\frac{3+\lambda_0}{4},\frac{9+\lambda_0 }{4};\frac{2+ \lambda_0}{2}; x\right)\Bigg]dr\nb\\
&& + \varphi_0, 
 \eqn
where $b_{1,2}$ are constants, and $\lambda_0$ is given by Eq.(\ref{4.7h}). Substituting the above into Eq.(\ref{D.3}), we find that 
 \bqn
  \label{7.10}
A(r)=\sqrt{1-\frac{2B}{r}}\left(A_0-
\int{\frac{\hat{P}(r)}{\sqrt{1-\frac{2B}{r}}}dr}\right),
 \eqn
with
 \bqn
 \label{7.11}
\hat{P}(r)&=&\frac{1}{4(2B-r)r^2}\Bigg\{\Big[(21-33\lambda)B^2+2(3-4\lambda)r^2\nb\\
&&+2Br(16\lambda-11)\Big](\varphi')^2-4Br\nb\\
&&-r^2(r-2B)^2(\lambda-1)(\varphi'')^2\Bigg\}\nb\\
&&+\frac{\lambda-1}{2r}\varphi'\left[(2B-r)r\varphi'''-2B\varphi''\right].
 \eqn
 
When $b_1 = b_2 = 0$, the above solutions reduce to
\bq
\lb{7.12}
\varphi(r) = \varphi_0,\;\;\;
A = 1 - A_0\sqrt{1-\frac{2B}{r}}.
\eq
Substituting it into the metric (\ref{id}), and considering the gauge
choice $h=0$, we find that it takes exactly the form of 
Eq.(\ref{7.5}) with the metric coefficients given by Eq.(\ref{7.6}) and
with the replacement $\varphi_1\to -2A_0$. Thus, the PPN parameters
$\beta$ and $\gamma$ are given by (\ref{7.8}) with $\varphi_1$ replaced
by $-2A_0$. For $A_0 \simeq {\cal{O}}(1)$, we obtain the 
constraint $|\upsilon-1|< {\cal{O}}\left(10^{-5}\right)$ again from
(\ref{6.2}). Therefore, the prescription (\ref{id}) leads to consistent 
results with solar system tests even for $\lambda \not= 1$.

\section{Conclusions}
\renewcommand{\theequation}{7.\arabic{equation}} \setcounter{equation}{0}

In this paper, we have studied spherically symmetric, stationary vacuum
configurations in the general covariant theory of the
Ho\v{r}ava-Lifshitz gravity with the projectability condition $N=N(t)$,
and an arbitrary value of the coupling constant $\lambda$
\cite{HMT,Silva,WW,HW}. In particular, in Sec. III we have obtained all
the solutions with the assumed symmetry in closed forms.

When applying these solutions to the solar system tests (Sec. IV), we
have shown explicitly that the ADM-type identification (\ref{idd})
between the metric coefficients and the basic quantities $N, N^i$ and 
$g_{ij}$ do not render the $\lambda\not=1$ solutions consistent 
with solar system tests, no matter how small $|\lambda-1|$ is. (On the other
hand, when $\lambda=1$ exactly, there is a spherically-symmetric,
stationary vacuum solution which is consistent with the solar system
tests \cite{GSW}.)

To show that this is indeed the case in more general situations, we have 
devoted Sec. V to consider the most general ansatz (\ref{metricB}) and
(\ref{metricC}) with spherical symmetry and shown that the prescription
(\ref{idd}) never recovers the Schwarzschild geometry in the 
$\lambda\to 1$ limit with $\Lambda_g=0$. Thus, one needs either to
invent a mechanism to restrict $\lambda$ precisely to its relativistic
value $\lambda_{GR}=1$, or to consider the gauge field $A$ and/or the
Newtonian prepotential $\varphi$ as parts of the $4$-dimensional metric
on which matter fields propagate.

In the case $\lambda = 1$, HMT proposed the identification (\ref{6.3a})
\cite{HMT} but clearly it is not gauge-invariant. Requiring the line
element be gauge-invariant not only under ${\mbox{Diff}}(M, {\cal{F}})$ 
(\ref{1.2}),  but also under the enlarged symmetry (\ref{symmetry}),  in
Sec. VI we have proposed the identification (\ref{id}), where $\upsilon$
is a dimensionless constant to be constrained by
observations/experiments. When $\upsilon=1$, it reduces to (\ref{6.3a})
in the gauge $\varphi = 0$.  Applying such a prescription to the cases
both with $\lambda = 1$ and with $\lambda \not= 1$, we have shown
that the resulted metric is indeed consistent with the solar system
tests, provided that $|\upsilon-1|<10^{-5}$. With such identifications,
one can also see the physical and geometrical roles that $A$ and
$\varphi$ play. In particular, the Newtonian prepotential $\varphi$ is
tightly related to the shift vector, while the geometrical lapse
function ${\cal{N}}$ is related to both $A$ and $\varphi$.

Finally, we note that it still remains to be understood how to obtain
the prescription (\ref{id}) (with $\upsilon\simeq 1$) from the 
action principle \footnote{In \cite{AdaSilva} the coupling of the HL
covariant theory with matter was considered from the action
principle. It was shown that Newtonian gravity cannot be recovered in
the weak gravitational field approximation, based on several
assumptions, including the one that the coupling among matter, the gauge
field $A$ and the Newtonian prepotential $\varphi$ be described by the
recipe provided in \cite{Silva}.}. 
Actually, in the UV, $N$ and $A-{\cal{A}}$ have different scaling
dimensions and thus, it is not easy to imagine how their linear
combination can universally enter the UV action of matter fields. On the
other hand, in the IR, $N$ and $A-{\cal{A}}$ have the same scaling
dimensions (they are actually dimensionless) and thus, the prescription
(\ref{id}) is not forbidden a priori. It is therefore important to
investigate whether the prescription (\ref{id}) can emerge in the IR
and, if it does, how.

~\\{\bf Acknowledgments:}   A.W. was supported in part by DOE  Grant,
DE-FG02-10ER41692. K.L. was partially supported by NSFC No. 11178018 and
No. 11075224.  S.M. was supported by the World Premier International
Research Center Initiative (WPI Initiative), MEXT, Japan;  Grant-in-Aid
for Scientific Research 24540256 and 21111006, and by Japan-Russia
Research Cooperative Program.

 \section*{Appendix A: Field Equations }
\renewcommand{\theequation}{A.\arabic{equation}} \setcounter{equation}{0}

Corresponding to the actions (\ref{2.4}), the  Hamiltonian and momentum constraints are given respectively by,
 \bqn
 \lb{eq1}
& & \int{ d^{3}x\sqrt{g}\left[{\cal{L}}_{K} + {\cal{L}}_{{V}} - \varphi {\cal{G}}^{ij}\nabla_{i}\nabla_{j}\varphi
- \big(1-\lambda\big)\big(\nabla^{2}\varphi\big)^{2}\right]}\nb\\
& & ~~~~~~~~~~~~~~~~~~~~~~~~~~~~~
= 8\pi G \int d^{3}x {\sqrt{g}\, J^{t}},\\
\lb{eq2}
& & \nabla^{j}\Big[\pi_{ij} - \varphi  {\cal{G}}_{ij} - \big(1-\lambda\big)g_{ij}\nabla^{2}\varphi \Big] = 8\pi G J_{i},
 \eqn
where %$J^{i} = g^{ij}J_{i}$, and
 \bqn
  \lb{eq2b}
  J^{t} &\equiv& 2 \frac{\delta\left(N{\cal{L}}_{M}\right)}{\delta N},\nb\\
   \pi_{ij} &\equiv& %\frac{\delta{\cal{L}}_{K}}{\delta\dot{g}_{ij}} =
   - K_{ij} +  \lambda K g_{ij},\nb\\
 J_{i} &\equiv& - N\frac{\delta{\cal{L}}_{M}}{\delta N^{i}}.
 \eqn

Variation of the action (\ref{2.4}) with respect to   $\varphi$ and $A$ yield, respectively,
\bqn
\lb{eq4a}
& & {\cal{G}}^{ij} \Big(K_{ij} + \nabla_{i}\nabla_{j}\varphi\Big) + \big(1-\lambda\big)\nabla^{2}\Big(K + \nabla^{2}\varphi\Big) \nb\\
& & ~~~~~~~~~~~~~~~~~~~~~~~~~~~~~~~~ = 8\pi G J_{\varphi}, \\
\lb{eq4b}
& & R - 2\Lambda_{g} =   8\pi G J_{A},
\eqn
where
\bq
\lb{eq5}
J_{\varphi} \equiv - \frac{\delta{\cal{L}}_{M}}{\delta\varphi},\;\;\;
J_{A} \equiv 2 \frac{\delta\left(N{\cal{L}}_{M}\right)}{\delta{A}}.
\eq
On the other hand,  the dynamical equations now read,
 \bqn \lb{eq3}
&&
\frac{1}{N\sqrt{g}}\Bigg\{\sqrt{g}\Big[\pi^{ij} - \varphi {\cal{G}}^{ij} - \big(1-\lambda\big) g^{ij} \nabla^{2}\varphi\Big]\Bigg\}_{,t} %^{\displaystyle{\cdot}}
\nb\\
& &~~~ = -2\left(K^{2}\right)^{ij}+2\lambda K K^{ij} \nb\\
& &  ~~~~~ + \frac{1}{N}\nabla_{k}\left[N^k \pi^{ij}-2\pi^{k(i}N^{j)}\right]\nb\\
& &  ~~~~~ - 2\big(1-\lambda\big) \Big[\big(K + \nabla^{2}\varphi\big)\nabla^{i}\nabla^{j}\varphi + K^{ij}\nabla^{2}\varphi\Big]\nb\\
& & ~~~~~ + \big(1-\lambda\big) \Big[2\nabla^{(i}F^{j)}_{\varphi} - g^{ij}\nabla_{k}F^{k}_{\varphi}\Big]\nb\\
& & ~~~~~ +  \frac{1}{2} \Big({\cal{L}}_{K} + {\cal{L}}_{\varphi} + {\cal{L}}_{A} + {\cal{L}}_{\lambda}\Big) g^{ij} \nb\\
& &  ~~~~~    + F^{ij} + F_{\varphi}^{ij} +  F_{A}^{ij} + 8\pi G \tau^{ij},
 \eqn
where $\left(K^{2}\right)^{ij} \equiv K^{il}K_{l}^{j},\; f_{(ij)}
\equiv \left(f_{ij} + f_{ji}\right)/2$, and
 \bqn
\lb{eq3a}
F^{ij} &\equiv&
\frac{1}{\sqrt{g}}\frac{\delta\left(-\sqrt{g}
{\cal{L}}_{V}\right)}{\delta{g}_{ij}}
 = \sum^{8}_{s=0}{g_{s} \zeta^{n_{s}}
 \left(F_{s}\right)^{ij} },\nb\\
F_{\varphi}^{ij} &=&  \sum^{3}_{n=1}{F_{(\varphi, n)}^{ij}},\nb\\
F_{\varphi}^{i} &=&  \Big(K + \nabla^{2}\varphi\Big)\nabla^{i}\varphi + \frac{N^{i}}{N} \nabla^{2}\varphi, \nb\\
F_{A}^{ij} &=& \frac{1}{N}\left[AR^{ij} - \Big(\nabla^{i}\nabla^{j} - g^{ij}\nabla^{2}\Big)A\right],\nb\\
 \eqn
with %The constants are given by $g_{0} = {2\Lambda}{\zeta^{-2}}$, $g_{1} = -1$, and
$n_{s} =(2, 0, -2, -2, -4, -4, -4, -4,-4)$. The
stress 3-tensor $\tau^{ij}$ is defined as
 \bq \label{tau}
\tau^{ij} = {2\over \sqrt{g}}{\delta \left(\sqrt{g}
 {\cal{L}}_{M}\right)\over \delta{g}_{ij}},
 \eq
and the geometric 3-tensors $ \left(F_{s}\right)_{ij}$ and $F_{(\varphi, n)}^{ij}$ are  given in \cite{HWWb}.

The matter components $(J^{t}, \; J^{i},\; J_{\varphi},\; J_{A},\; \tau^{ij})$ satisfy the
conservation laws,
 \bqn \lb{eq5a} & &
 \int d^{3}x \sqrt{g} { \left[ \dot{g}_{kl}\tau^{kl} -
 \frac{1}{\sqrt{g}}\left(\sqrt{g}J^{t}\right)_{, t}
 +   \frac{2N_{k}}  {N\sqrt{g}}\left(\sqrt{g}J^{k}\right)_{,t}
  \right.  }   \nb\\
 & &  ~~~~~~~~~~~~~~ \left.   - 2\dot{\varphi}J_{\varphi} -  \frac{A} {N\sqrt{g}}\left(\sqrt{g}J_{A}\right)_{,t}
 \right] = 0,\\
\lb{eq5b} & & \nabla^{k}\tau_{ik} -
\frac{1}{N\sqrt{g}}\left(\sqrt{g}J_{i}\right)_{,t}  - \frac{J^{k}}{N}\left(\nabla_{k}N_{i}
- \nabla_{i}N_{k}\right)   \nb\\
& & \;\;\;\;\;\;\;\;\;\;\;- \frac{N_{i}}{N}\nabla_{k}J^{k} + J_{\varphi} \nabla_{i}\varphi - \frac{J_{A}}{2N} \nabla_{i}A
 = 0.
\eqn

\section*{Appendix B: $G$ and $H$ defined in Eqs.(\ref{equ4}) and (\ref{equ4a}) }
\renewcommand{\theequation}{B.\arabic{equation}} \setcounter{equation}{0}

The functions $G$ and $H$ defined in Eqs.(\ref{equ4}) and (\ref{equ4a}) are given by
\bqn
\lb{B.1}
G(r) &=& rh \big[2h'-r (\lambda -1)h''\big]  +\frac{1}{2}r^2(\lambda -1)h'^2 \nb\\
&&   +\frac{r}{2f} \Big[r (\lambda -1) f''  -2 (\lambda +1) f'\Big]h^{2}\nb\\
&&  -\frac{3r^{2} f'^2}{8f^{2}} (\lambda -1) h^2 + (4 \lambda -3) h^2\nb\\
&&-\frac{1}{4\zeta^4 r^{4}} \Bigg\{8 (46 g_4+17 g_5+7 g_6+28 g_7+9 g_8)
f^3\nb\\
&&-4 \big[8g_7f^{(4)} r^4+3 g_8 f^{(4)} r^4+14 g_2 \zeta ^2 r^2+5
g_3
\zeta^2 r^2\nb\\
&&+(48 g_4+14 g_5+3 g_6-48 g_7-16g_8) f'' r^2\nb\\
&&-2 (48 g_4+14 g_5+3 g_6-48 g_7-16 g_8) f' r+180 g_4\nb\\
&&+66 g_5+27 g_6+48 g_7+12g_8\big] f^2+\big[-3 (8 g_7\nb\\
&&+3 g_8) (f'')^2 r^4+(48 g_4+22 g_5+12 g_6+48 g_7\nb\\
&&+13 g_8) (f')^2 r^2+4 (8g_2 r^2 \zeta ^2+3 g_3 r^2 \zeta ^2+48
g_4\nb\\
&&+14 g_5+3 g_6-16 g_7-4 g_8) f'' r^2-2 f' \big[3 (32 g_4\nb\\
&&+12g_5+5 g_6-g_8) f'' r^2+2 \big((8 g_7+3 g_8) f''' r^3\nb\\
&&+96 g_4+28 g_5+6 g_6-32 g_7-8g_8\big)\big] r+4 \big[r^4 \zeta ^4\nb\\
&&+12 g_2 r^2 \zeta ^2+4 g_3 r^2 \zeta ^2+84 g_4+30 g_5+12 g_6-8
g_7\nb\\
&&-6g_8\big]\big] f+(32 g_4+12 g_5+5 g_6-g_8) r^3 (f')^3\nb\\
&&+4 \big[\zeta ^4 \Lambda r^6-\zeta ^4 r^4+2 g_2 \zeta ^2 r^2+g_3
\zeta ^2 r^2+4 g_4\nb\\
&&+2 g_5+g_6\big]+r^2 (f')^2 \big[-8 g_2 \zeta ^2 r^2-3 g_3 \zeta^2
r^2\nb\\
&&+(8 g_7+3 g_8) f'' r^2-48 g_4-14 g_5\nb\\
&&-3 g_6+16 g_7+4g_8\big]\Bigg\},\nb\\
 H(r) &=& 2\lambda r^2 h h'' + \big(\lambda + 1\big)r^{2} \left(h'\right)^{2} +  4\left(2\lambda-1\right)r hh'\nb\\
 &&  - \frac{(2\lambda + 1)r^{2}f'}{f} h h' -  \frac{r}{f}  \Big[\lambda r f'' + (2\lambda - 1) f'\Big] h^{2}\nb\\
&&  + \frac{r^{2}}{4f^{2}}  (5 \lambda +1) (f')^2 h^2\nb\\
&&+\frac{1}{4 r^4 \zeta ^4}\Bigg\{32 (46 g_4+17 g_5+7 g_6+28 g_7+9
g_8)f^3\nb\\
&&+4 \big[8 g_7 f^{(5)} r^5+3g_8 f^{(5)} r^5+48 g_4 f^{'''} r^3\nb\\
&&+14 g_5 f^{'''}r^3+3 g_6 f^{'''} r^3-48g_7 f^{'''} r^3\nb\\
&&-16 g_8 f^{'''} r^3-28g_2 \zeta ^2 r^2-10g_3\zeta ^2 r^2-4
(48g_4\nb\\
&&+14g_5+3g_6-48g_7-16g_8) f'' r^2+6 (2g_4-3g_5\nb\\
&&-4 g_6-76 g_7-25 g_8) f'
r-720g_4-264g_5\nb\\
&&-108g_6-192g_7-48g_8\big]f^2+2 \big[-16g_2 \zeta ^2 f^{'''}
r^5\nb\\
&&-6g_3\zeta ^2 f^{'''} r^5+3(32 g_4+12g_5+5g_6-g_8) (f'')^2
r^4\nb\\
&&-96g_4 f^{'''}r^3-28g_5 f^{'''} r^3-6 g_6f^{'''} r^3+32 g_7f^{'''}
r^3\nb\\
&&+8 g_8f^{'''} r^3+48g_2\zeta^2 r^2+16g_3\zeta^2 r^2-3
(112g_4\nb\\
&&+30g_5+4g_6-144g_7-47g_8) (f')^2 r^2+f'' \big(5 (8 g_7\nb\\
&&+3 g_8) f^{'''} r^3+8 (48 g_4+14 g_5+3 g_6-16 g_7\nb\\
&&-4 g_8)\big) r^2+f' \bigg(3 (32g_4+12g_5+5g_6-g_8) f^{'''}
r^3\nb\\
&&+(48g_4-2g_5-15g_6-240g_7-74g_8) f'' r^2\nb\\
&&+2 \big(3 (8g_7+3g_8) f^{(4)} r^4+2 (14g_2r^2 \zeta ^2+5g_3r^2
\zeta^2\nb\\
&&+36g_4+24 g_5+18 g_6+96g_7+24g_8)\big)\bigg) r+672g_4\nb\\
&&+240 g_5+96 g_6-64g_7-48 g_8\big]f-(16g_4+10g_5\nb\\
&&+7g_6+48g_7+14g_8) r^3 (f')^3+8(-\zeta ^4 \Lambda  r^6+2g_2\zeta^2
r^2\nb\\
&&+g_3\zeta^2 r^2+8g_4+4g_5+2g_6)-r f' \big[-(8g_7\nb\\
&&+3g_8) (f'')^2 r^4+2 (8g_2r^2 \zeta^2+3g_3r^2 \zeta^2+48
g_4\nb\\
&&+14g_5+3g_6-16g_7-4g_8) f'' r^2+4 (r^4 \zeta^4\nb\\
&&+12g_2r^2 \zeta^2+4 g_3r^2
\zeta^2+84g_4+30g_5+12g_6\nb\\
&&-8g_7-6g_8)\big]+3 r^2 (f')^2 \big[(8g_7+3g_8) f^{'''}
r^3\nb\\
&&+(32g_4+12g_5+5g_6-g_8) f'' r^2+96 g_4+28g_5\nb\\
&&+6 g_6-32g_7-8g_8\big]\Bigg\},
\eqn
where $f^{(n)} \equiv d^{n}f/dr^{n}$.

\section*{Appendix C: The U(1) Gauge Transformations and Gauge Choices}
\renewcommand{\theequation}{C.\arabic{equation}} \setcounter{equation}{0}

Under the U(1) gauge transformations (\ref{Ugauge}), in the spherically symmetric case, the variables $(N, N^i, g_{ij}, A, \varphi)$ transform as,
\bqn
\lb{UgaugeA}
\delta_{\alpha}{N} &=& 0,\;\;\;
\delta_{\alpha}N^{i}  = \delta_{r}^{i}f\alpha',\;\;\;
\delta_{\alpha}g_{ij} = 0,\nb\\
\delta_{\alpha}A &=&\dot{\alpha} - h\alpha',\;\;\;
\delta_{\alpha}\varphi = - \alpha,
\eqn
 where  $\alpha = \alpha(t, r)$. From these expressions, one can see that various gauges  can be chosen.

 \subsection{ $\varphi = 0$}

 In this gauge, we have
 \bq
 \lb{gaugeA}
 \alpha = \varphi(t,r),
 \eq
 which is unique, and is the gauge used in Section III.

 \subsection{  $A = 0$}

 In this gauge, we have
 \bq
 \lb{gaugeB}
\dot{\alpha} - h \alpha' = - A.
 \eq

When $h = 0$, we have
\bq
\lb{C.1}
\alpha(t, r) = - \int^{t}{A(t', r) dt'} + \alpha_{0}(r),
\eq
where  $\alpha_{0}(r)$ is an arbitrary function of its indicated argument. Thus, in this case the gauge is fixed only up to  an
arbitrary function of $r$.

When $h \not= 0$, we introduce two new variables $u$ and $v$ via the relations,
\bqn
\lb{C.1a}
dt &=& G dv + F du ,\nb\\
dr &=&  h(G dv-Fdu),
\eqn
where $F$ and $G$ are functions of $u$ and $v$ only, and satisfy the integrability conditions,
\bqn
\lb{C.1b}
F_{,v} - G_{,u}  &=& 0,\\
\lb{C.1ba}
(Fh)_{,v} + (Gh)_{,u} &=& 0.
\eqn
Note that one should not consider Eq.(\ref{C.1a}) as coordinate transformations, because they are forbidden by ${\mbox{Diff}}(M, {\cal{F}})$, but rather a technique to solve
Eq.(\ref{gaugeB}). Then, in terms of $u$ and $v$, Eq.(\ref{gaugeB}) takes the form, $\alpha_{,u} = - FA$, which has the solution,
\bq
\lb{C.1c}
\alpha(t, r)  = - \int^{u}{F(u', v) A(u', v) du'} + \alpha_{1}(v),
\eq
where $\alpha_{1}$ is an arbitrary function of $v$ only, and $u = u(t, r)$ and $v = v(t, r)$, given through Eqs.(\ref{C.1a})-(\ref{C.1ba}). Therefore, 
in the present case the gauge is fixed up to an arbitrary function of
$v$.

\subsection{$h = 0$}

In this gauge, we have
\bq
\lb{C.3}
\alpha' = - \frac{h}{f},
\eq
which has the solution,
\bq
\lb{C.4}
\alpha(t, r) = - \int^{r}{\frac{h(t,r')dr'}{f(t,r')} } + \alpha_{2}(t),
\eq
where $\alpha_{2}(t)$ is an arbitrary function of $t$ only.

\section*{Appendix D: Field Equations without Specifying the U(1) Gauge }
\renewcommand{\theequation}{D.\arabic{equation}} \setcounter{equation}{0}

It can be shown that in the spherically symmetric case, there are only three independent field equations: the constraint obtained from the variation of the gauge field $A$ given by Eq.(\ref{eq4a}),
the momentum constraint (\ref{eq2}), and the rr-componet of the dynamical equations (\ref{eq3}). For the ADM decomposition given by Eq.(\ref{4.1}), they read, respectively, 
\bqn 
\lb{D.1}
&&  (rf)'  - \left(1 - \Lambda_gr^2\right) = 0,\\
\lb{D.2}
&& (1-\lambda)\Bigg\{r^2f^2h''-\frac{rf}{2}(rf'-4f)h'-\Big[2f^2\nb\\
&& -\frac{r^2}{2}(f')^2+\frac{r^2}{2}ff''\Big]h-\frac{f^2}{2}\Big[4f\varphi'-4rf'\nb\\
&& -r^2f''-(4rf+3r^2f')\varphi''-2rf\varphi'''\Big]\Bigg\}\nb\\
&& -rff'h+f^2(f-1+\Lambda_gr^2)\varphi' = 0,\\
\lb{D.3}
%\Big\{
&& \frac{16}{r}A'+\frac{8}{r^2f}(f-1+r^2\Lambda_g)A\nb\\
&& +\frac{4}{r^2}(3f-1+r^2\Lambda_g)\varphi'(\varphi'+h)+\frac{16hh'}{rf}\nb\\
&& +\frac{8h^2}{r^2f^2}(f-2rf')-\frac{2}{r^6\zeta^4f}\Bigg[8 (46 g_4+17
g_5+7 g_6\nb\\
&& -28 g_7+9 g_8) f^3-4 \Big(-(8 g_7-3 g_8) f^{(4)}r^4+(14 g_2\nb\\
&& +5 g_3) \zeta^2 r^2+(48 g_4+14 g_5+3 g_6+48g_7\nb\\
&& -16 g_8) (r f''-2 f') r+180 g_4+66 g_5+27 g_6\nb\\
&& +12 (g_8-4g_7)\Big) f^2+\Big(4 \big(r^4 \zeta ^4+4 (3g_2+g_3) r^2
\zeta^2\nb\\
&& +84 g_4+30 g_5+12 g_6+8 g_7-6 g_8\big)+r\big((48 g_4\nb\\
&& +22g_5+12 g_6-48 g_7+13 g_8) r (f')^2-2 \big(2 (3 g_8\nb\\
&& -8g_7) f''' r^3+3 (32 g_4+12g_5+5 g_6-g_8) f'' r^2\nb\\
&& +4 (48g_4+14g_5+3 g_6+16 g_7-4 g_8)\big) f'\nb\\
&& +r f''\big(3 (8 g_7-3g_8) f'' r^2+4 ((8 g_2+3 g_3) r^2 \zeta
^2+48g_4\nb\\
&& +14 g_5+3 g_6+16 g_7-4 g_8)\big)\big)\Big)f+4 \Big(r^2 \big(r^2 (r^2
\Lambda -1) \zeta^2\nb\\
&& +2 g_2+g_3\big) \zeta ^2+4 g_4+2 g_5+g_6\Big)-r^2 (f')^2
\big((8g_2\nb\\
&& +3 g_3) r^2 \zeta^2+48 g_4+14 g_5+3 g_6+16 g_7-4 g_8\nb\\
&& +r((-32g_4-12 g_5-5 g_6+g_8)f'+(8 g_7-3 g_8) r
f'')\big)\Bigg]\nb\\
&& + (1-\lambda) \Bigg\{\bigg[4f''-\frac{(f')^2}{f}+\frac{8f'}{r}-\frac{32f}{r^2}\bigg]\varphi^2+8f'\varphi'\varphi''\nb\\
&& +\frac{h^2}{r^2f^3}\bigg[3r^2(f')^2-32f^2-4rf(rf''-2f')\bigg]\nb\\
&& \frac{8}{f}hh''+\left(\frac{16f'}{rf}-\frac{64}{r^2}+\frac{6(f')^2}{f^2}\right)\varphi'h\nb\\
&& +\frac{8f'}{f}(2\varphi''h-\varphi'h')+8\varphi'h''-\frac{4}{f}(h'+f\varphi'')^2\nb\\
&& +8(h+f\varphi')\varphi'''\Bigg\}=0.
 \eqn

%It should be noted that the metric (\ref{smetric}) is gauge-invariant, so it does not depend on the choices of these specific gauges.

%%%%%%%%%%%%%%%%%%%%%%%%%%%%%%%%%%%%%%%%%%%%%%%%%%%%%%


\begin{thebibliography}{nbound}


\bibitem{Will05} C.M. Will, Living Rev. Relativity, {\bf 9}, 3 (2006);
E. Komatsu, {\em et al},   Astrophys. J. Suppl. {\bf 192}, 18 (2011) [arXiv:1001.4538].


\bibitem{HV} G. 't Hooft, Nucl. Phys. B{\bf 62}, 444 (1973); G. 't Hooft and M. Veltman, Ann. Inst.
Poincare, {\bf 20},  69  (1974); S. Deser, P. Van Nieuwenhuizen, Phys. Rev. D{\bf 10},  401  (1974); 411 (1974).

\bibitem{QGs} S. Weinberg, in {\em General Relativity, An Einstein Centenary Survey}, edited by S.W. Hawking and W. Israel (Cambridge University Press, Cambridge, 1980);
      C. Kiefer, {\em Quantum Gravity} (Oxford Science Publications, Oxford University Press, 2007);
      K. Becker, M. Becker, and J.H. Schwarz, {\em String Theory and M-Theory} (Cambridge University Press, Cambridge, 2007);
      C. Rovelli,   {\em Quantum gravity} (Cambridge University Press, Cambridge, 2008).


\bibitem{Horava} P. Horava, JHEP, {\bf 0903}, 020 (2009) [arXiv:0812.4287]; Phys.
Rev. D{\bf 79}, 084008 (2009) [arXiv:0901.3775]; and Phys. Rev.
Lett. {\bf 102}, 161301  (2009) [arXiv:0902.3657].

\bibitem{Stelle}  K.S. Stelle, Phys. Rev. D{\bf 16},   953 (1977).

\bibitem{LZbreaking}  D. Mattingly, Living Rev.  Relativity, {\bf 8}, 5 (2005); S. Liberati and L. Maccione, Annu. Rev. Nucl. Part. Sci. {\bf 59}, 245 (2009).

\bibitem{Pola}  J. Polchinski, arXiv:1106.6346.


\bibitem{Lifshitz} E.M. Lifshitz, Zh. Eksp. Teor. Fiz. {\bf 11}, 255; 269 (1941).

\bibitem{Visser} M. Visser, Phys. Rev. D{\bf 80}, 025011 (2009) [arXiv:0902.0590];   arXiv:0912.4757.


\bibitem{ADM}  C.W. Misner, K.S. Thorne, and J.A. Wheeler, {\em Gravitation } (W.H. Freeman and Company, San Francisco, 1973),
pp.484-528.



  \bibitem{BPS}     D. Blas,  O. Pujolas, and S. Sibiryakov,  Phys. Rev. Lett. {\bf 104}, 181302 (2010) [arXiv:0909.3525];
                          JHEP, {\bf 1104}, 018 (2011) [arXiv.1007.3503].


 \bibitem{KP}      I. Kimpton and A. Padilla, J. High Energy Phys. {\bf 07}, 014 (2010) [arXiv:1003.5666].


  \bibitem{ZSWW} T.  Zhu, F.-W.  Shu, Q. Wu, and A. Wang,  Phys. Rev. D{\bf 85}, 044053 (2012) [arXiv:1110.5106].


 \bibitem{BLW} A.  Borzou, K. Lin, and A. Wang,  J. Cosmol. Astropart. Phys., {\bf 05}, 006 (2011) [arXiv:1103.4366].


  \bibitem{AdSCFT}   J.M. Maldacena,  Adv. Theor. Math. Phys.  {\bf 2},  231 (1998); % [arXiv:hep-th/9711200].
  %;
  O. Aharony, S.S. Gubser, J. Maldacena,   H. Ooguri, and Y. Oz,  Phys. Rept.  {\bf 323}, 183 (2000).


  \bibitem{OM} L. Onsager and S. Machlup, Phys. Rev. {\bf 91}, 1505 (1953); S. Machlup and L. Onsager, {\em ibid.}, {\bf 91}, 1512 (1953).


  \bibitem{Verlinde}  E.P. Verlinde,  JHEP,  {\bf 04}, 029 (2011). %  [arXiv:1001.0785].



    \bibitem{Hreview}        P. Horava, Class. Quantum Grav. {\bf 28},   114012 (2011) [arXiv:1101.1081].

  %\cite{Mukohyama:2010xz}
\bibitem{Mukohyama:2010xz}
  S.~Mukohyama,
 % ``Horava-Lifshitz Cosmology: A Review,''
  Class.\ Quant.\ Grav.\  {\bf 27}, 223101 (2010)
  [arXiv:1007.5199]. % [hep-th]].
  %%CITATION = CQGRD,27,223101;%%


\bibitem{Calcagni} G. Calcagni,  JHEP, {\bf 09}, 112 (2009) [arXiv:0904.0829];
  R. Brandenberger, Phys. Rev. D{\bf 80}, 043516 (2009) [arXiv:0904.2835];
                            A. Wang and Y. Wu, JCAP, {\bf 07}, 012 (2009) [arXiv:0905.4117];
  Y.~Misonoh, K.~-i.~Maeda, T.~Kobayashi,  [arXiv:1104.3978].


  \bibitem {KK} E. Kiritsis and G. Kofinas,  Nucl. Phys. B{\bf 821}, 467 (2009) [arXiv:0904.1334].



 %\cite{Mukohyama:2009tp}
\bibitem{Mukohyama:2009tp}
  S.~Mukohyama,
  %``Caustic avoidance in Horava-Lifshitz gravity,''
  JCAP {\bf 0909}, 005 (2009).
  [arXiv:0906.5069].




 %\cite{Mukohyama:2009gg}
\bibitem{Mukohyama:2009gg}
  S.~Mukohyama,
  %``Scale-invariant cosmological perturbations from Horava-Lifshitz gravity without inflation,''
  JCAP {\bf 0906}, 001 (2009).
  [arXiv:0904.2190].




 \bibitem {MMS} S. Maeda, S. Mukohyama and T. Shiromizu, Phys. Rev. D{\bf 80}, 123538 (2009) [arXiv:0909.2149]. %[astro-ph.CO]].

  \bibitem {MNTY} S. Mukohyama, K. Nakayama, F. Takahashi and S. Yokoyama, Phys. Lett. B{\bf 679}, 6 (2009) [arXiv:0905.0055]. % [hep-th]].


   \bibitem {Mukohyama:2009mz} S. Mukohyama,    Phys. Rev. D{\bf 80}, 064005 (2009) [arXiv:0905.3563];


   \bibitem {Wang}    A. Wang,  Mod. Phys. Lett. A{\bf 26}, 387 (2011) [arXiv:1003.5152].

   \bibitem {TS}  T. Takahashi and J. Soda, Phys. Rev. Lett. {\bf 102}, 231301 (2009) [arXiv:0904.0554]. % [hep-th]].





     \bibitem{Ins}       A. Wang and R. Maartens,  Phys. Rev. D{\bf 81}, 024009 (2010)  [arXiv:0907.1748].



  \bibitem{KA}  K. Koyama and F. Arroja, JHEP,   {\bf 03}, 061 (2010)  [arXiv:0910.1998].



\bibitem{WWa} A. Wang and Q. Wu, Phys. Rev. D{\bf 83}, 044025 (2011)  [arXiv:1009.0268].


\bibitem{SC}   C. Charmousis, G. Niz, A. Padilla, and P.M. Saffin, JHEP,  {\bf 08}, 070 (2009) [arXiv:0905.2579];
                            D. Blas, O. Pujolas, and S. Sibiryakov, {\em ibid.},   {\bf 10}, 029 (2009)  [arXiv:0906.3046];
                            A. Papazoglou and T.P. Sotiriou, Phys. Lett. B{\bf 685}, 197 (2010)  [arXiv:0911.1299].



 \bibitem{ZWWS} T.  Zhu,  Q. Wu, A. Wang, and F.-W.  Shu,  Phys. Rev. D{\bf  84}, 101502(R) (2011)   [arXiv:1108.1237].
 
 
 \bibitem{reviews}  A. Padilla,   J. Phys. Conf. Ser. {\bf 259}, 012033 (2010)  [arXiv:1009.4074];
T.P. Sotiriou,  J. Phys. Conf. Ser. {\bf 283}, 012034 (2011)  [arXiv:1010.3218];
% P. Ho\v{r}ava, Class. Quantum Grav. {\bf 28},   114012 (2011)  [arXiv:1101.1081];
T. Clifton, P.G. Ferreira, A. Padilla,  and C. Skordis,  arXiv:1106.2476.



\bibitem{Izumi:2011eh}
  K.~Izumi and S.~Mukohyama,
  %``Nonlinear superhorizon perturbations in Horava-Lifshitz gravity,''
  Phys. Rev. D{\bf 84}, 064025
(2011) [arXiv:1105.0246]. % [hep-th].
  %%CITATION = ARXIV:1105.0246;%%


\bibitem{GSWa}  A. Emir G\"umr\"uk\c{c}\"uo\u{g}lu, S.~Mukohyama, and A. Wang, Phys. Rev. D{\bf  85}, 064042 (2012) [arXiv:1109.2609].

 
\bibitem{Vainshtein:1972sx}
  A.~I.~Vainshtein,
  %``To the problem of nonvanishing gravitation mass,''
  Phys.\ Lett.\  B {\bf 39}, 393 (1972).
  %%CITATION = PHLTA,B39,393;%%



 \bibitem{HMT} P. Horava and C.M. Melby-Thompson, Phys. Rev. D{\bf 82}, 064027 (2010) [arXiv:1007.2410].


    \bibitem{WW}  A. Wang and Y. Wu,   Phys. Rev. D{\bf 83}, 044031 (2011) [arXiv:1009.2089].



  \bibitem{Silva} A.M. da Silva, Class. Quantum Grav. {\bf 28}, 055011 (2011) [arXiv:1009.4885].

 \bibitem{HW} Y.-Q. Huang and A. Wang, Phys. Rev. D{\bf 83}, 104012   (2011) [arXiv:1011.0739].
 
 
 
  \bibitem{HWWb} Y.-Q. Huang, A. Wang, and Q. Wu,  arXiv:1201.4630.


\bibitem{SST} R.A. Konoplya,   Phys. Lett. B{\bf 679},  499  (2009) [arXiv:0905.1523];
                         T. Harko, Z. Kovacs and F.S.N. Lobo, Phys. Rev. D{\bf 80}, 044021 (2009) [arXiv:0907.1449]; 
                                           Class. Quant. Grav. {\bf 28}, 165001 (2011) [arXiv:1009.1958]; 
                                            Proc. Roy. Soc. Lond, A{\bf 467}, 1390 (2011) [arXiv:0908.2874];
                         L. Iorio and M.L. Ruggiero,  Int. J. Mod. Phys. A{\bf 25},   5399 (2010) [arXiv:0909.2562]; 
                                                   Open Astron. J., 2010, {\bf 3}, 167 (2010) [arXiv:0909.5355];
                                                   Inter. J. Mod. Phys. D{\bf 20}, 1079 (2011) [arXiv:1012.2822];
                        Z. Horvath, L.A. Gergely, Z. Keresztes, T. Harko, and F.S. N. Lobo, Phys. Rev. D{\bf 84}, 083006  (2011) [arXiv:1105.0765].
 

  \bibitem{GSW} J. J. Greenwald, V.H. Satheeshkumar, and A. Wang, JCAP, {\bf 12}, 007 (2010) [arXiv:1010.3794].
  
  
  
    \bibitem{GLLSW}  J. Greenwald, J. Lenells, J. X. Lu, V. H. Satheeshkumar, and A. Wang,  Phys. Rev. D{\bf 84}, 084040  (2011) [arXiv:1105.4259].   %


        \bibitem{BLWb} A.  Borzou, K. Lin, and A. Wang,   JCAP, {\bf 02}, 025 (2012) [arXiv:1110.1636].



 \bibitem{WM} A. Wang and R. Maartens,  Phys. Rev. D {\bf 81}, 024009 (2010) [arXiv:0907.1748];
 A. Wang, D. Wands, and R. Maartens,      J. Cosmol. Astropart. Phys.,  {\bf 03}, 013  (2010) [arXiv:0909.5167].



 \bibitem{SVW}  T. Sotiriou, M. Visser, and S. Weinfurtner,  Phys. Rev. Lett. {\bf 102},  251601 (2009)
                       [arXiv:0904.4464]; J. High Energy Phys.,  {\bf 10}, 033 (2009) [arXiv:0905.2798]. %;  arXiv:1002.0308.


   


 \bibitem{AP} J.  Alexandre and P. Pasipoularides,   Phys. Rev. D{\bf 83}, 084030 (2011) [arXiv:1010.3634].
 


    \bibitem{GPW} J. Greenwald, A. Papazoglou, and A. Wang, Phys. Rev. D{\bf 81}, 084046 (2010) [arXiv:0912.0011].
    
   \bibitem{IM}   	K. Izumi and S. Mukohyama, Phys. Rev. D{\bf 81}, 044008 (2010) [arXiv:0911.1814].
 
 \bibitem{AP2} J.  Alexandre and P. Pasipoularides,   Phys. Rev. D{\bf 84}, 084020 (2011) [arXiv:1108.1348].


 


 \bibitem{AS72}  M. Abramowitz and I.A. Stegun, Handbook of Mathematical
Functions (Dover Publications, INC., New York, 1972).




  \bibitem{Edd}  A.S. Eddington, The Mathematical Theory of Relativity (Cambridge University Press, Cambridge, 1957).




    \bibitem{BT} B. Bertotti, L. Iess and P. Tortora, Nature, {\bf 425}, 374 (2003).

 \bibitem{WTB} A. Fienga,  et al,    Celestial Mechanics and Dynamical Astronomy, 3, 363 (2011).

 \bibitem{RJ}  A. Hees, et al, arXiv:1110.0659.



 \bibitem{BD}  C. Brans and R. H. Dicke, Phys. Rev. {\bf 124}, 925 (1961).
 
 
  \bibitem{DInverno} R. D'Inverno, {\em Introducing Einstein's Relativity} (Clarendon Press, Oxford, 2003), pp.165-168.
  
  
   \bibitem{AdaSilva}  E. Abdalla,  and A.M. da Silva, Phys. Lett. B{\bf 707}, 311 (2012) [arXiv:1111.2224].


 



\end{thebibliography}
\end{document}